\documentclass[12pt, preprint, longabstract]{aastex}
\usepackage{ccaption,equation}
\usepackage[multidot]{grffile}
\usepackage{booktabs}
\newcommand\ltsima{$\; \buildrel <\over\sim \;$}
\newcommand\simlt{\lower.5ex\hbox{\ltsima}}
\newcommand\gtsima{$\; \buildrel >\over\sim \;$}
\newcommand\simgt{\lower.5ex\hbox{\gtsima}}

\newcommand\mearth {{M_\oplus}}

\def\Vec#1{\mbox{\boldmath $#1$}}

\shorttitle{Cold Neptune Analog Exoplanet}
\shortauthors{Sumi et al.}

\begin{document}

\title{
The First Neptune Analog or Super-Earth with Neptune-like Orbit: MOA-2013-BLG-605Lb}

\author{
T.~Sumi\altaffilmark{1,2},
A.~Udalski\altaffilmark{3,4},
D.P.~Bennett\altaffilmark{5,6,2},
A.~Gould\altaffilmark{7},  
R. Poleski\altaffilmark{3,4,7},
I.A.~Bond\altaffilmark{8,2}, 
J.Skowron\altaffilmark{3,4},
N.~Rattenbury\altaffilmark{9,2},
R. W.~Pogge\altaffilmark{7},
T.~Bensby\altaffilmark{10},
J.P.~Beaulieu\altaffilmark{11} ,
J.B.~Marquette\altaffilmark{11} ,
V.~Batista\altaffilmark{11} ,
S.~Brillant\altaffilmark{12}
\\
and\\
F.~Abe\altaffilmark{13}, 
Y.~Asakura\altaffilmark{13}, 
A.~Bhattacharya\altaffilmark{5},
M.~Donachie\altaffilmark{9},
M.~Freeman\altaffilmark{9},
A.~Fukui\altaffilmark{14},
Y.~Hirao\altaffilmark{1},
Y.~Itow\altaffilmark{13},
N.~Koshimoto\altaffilmark{1},
M.C.A.~Li\altaffilmark{9},
C.H.~Ling\altaffilmark{8},
K.~Masuda\altaffilmark{13},
Y.~Matsubara\altaffilmark{13},
Y.~Muraki\altaffilmark{13},
M.~Nagakane\altaffilmark{1},
K.~Ohnishi\altaffilmark{15},
H.~Oyokawa\altaffilmark{13}, 
To.~Saito\altaffilmark{16},
A.~Sharan\altaffilmark{9},
D.J.~Sullivan\altaffilmark{17},
D.~Suzuki\altaffilmark{5},
P.,J.~Tristram\altaffilmark{18},
A. Yonehara\altaffilmark{19},\\ 
(The MOA Collaboration) \\
M.K.~Szyma{\'n}ski\altaffilmark{3},
{K}.~Ulaczyk\altaffilmark{3},
S. Koz\l{}owski \altaffilmark{3},
{\L}.~Wyrzykowski\altaffilmark{3},
M.~Kubiak\altaffilmark{3},
P. Pietrukowicz\altaffilmark{3},
G. Pietrzy{\'n}ski\altaffilmark{3},
I.~Soszy{\'n}ski\altaffilmark{3},
\\
(The OGLE Collaboration)\\
and\\
C. Han\altaffilmark{20},
Y.-K. Jung\altaffilmark{20}, 
I.-G Shin\altaffilmark{20},
C-U. Lee\altaffilmark{21}
}

\altaffiltext{1}{Department of Earth and Space Science, Graduate School of Science, Osaka University, Toyonaka, Osaka 560-0043, Japan,\\
e-mail: {\tt sumi@ess.sci.osaka-u.ac.jp}}
\altaffiltext{2}
{Microlensing Observations in Astrophysics (MOA)}
\altaffiltext{3}
{Warsaw University Observatory, Al.~Ujazdowskie~4, 00-478~Warszawa,Poland; udalski, msz, mk, pietrzyn, soszynsk, szewczyk, kulaczyk@astrouw.edu.pl}
\altaffiltext{4}
{Optical Gravitational Lens Experiment (OGLE)}
\altaffiltext{5}
{Department of Physics, University of Notre Dame, Notre Dame, IN 46556, USA; bennett@nd.edu}
\altaffiltext{6}{Laboratory for Exoplanets and Stellar Astrophysics, NASA/Goddard Space Flight Center, Greenbelt, MD 20771, USA}
\altaffiltext{7}
{Department of Astronomy, Ohio State University, 140 W.\ 18th Ave., Columbus, OH 43210, USA}

\altaffiltext{8}
{Institute of Information and Mathematical Sciences, Massey University,
Private Bag 102-904, North Shore Mail Centre, Auckland, New Zealand;
i.a.bond,c.h.ling,w.sweatman@massey.ac.nz}
\altaffiltext{9}
{Department of Physics, University of Auckland, Private Bag 92019, Auckland, New Zealand; n.rattenbury,mli351,p.yock@auckland.ac.nz; asha583,mdon849@aucklanduni.ac.nz}
\altaffiltext{10}
{Lund Observatory, Department of Astronomy and Theoretical physics, Box 43, SE-221 00 Lund, Sweden}
\altaffiltext{11}{UPMC-CNRS, UMR7095, Institut d'Astrophysique de Paris, F-75014 Paris, France}
\altaffiltext{12}{European Southern Observatory (ESO), Karl-Schwarzschildst. 2, D-85748 Garching, Germany}

\altaffiltext{13}{ Institute for Space-Earth Environmental Research, Nagoya University, Nagoya 464-8601, Japan; abe,itow,kmasuda,ymatsu@stelab.nagoya-u.ac.jp}
\altaffiltext{14}{Okayama Astrophysical Observatory, National Astronomical Observatory of Japan, 3037-5 Honjo, Kamogata, Asakuchi, Okayama 719-0232, Japan}
\altaffiltext{15}{Nagano National College of Technology, Nagano 381-8550, Japan}
\altaffiltext{16}{Tokyo Metropolitan College of Aeronautics, Tokyo 116-8523, Japan}
\altaffiltext{17}{School of Chemical and Physical Sciences, Victoria University, Wellington, New Zealand}
\altaffiltext{18}{Mt. John University Observatory, P.O. Box 56, Lake Tekapo 8770, New Zealand}
\altaffiltext{19}{Department of Physics, Faculty of Science, Kyoto Sangyo University, 603-8555 Kyoto, Japan}
\altaffiltext{20}{Department of Physics, Institute for Astrophysics, Chungbuk National University, 371-763 Cheongju, Korea}
\altaffiltext{21}{Korea Astronomy and Space Science Institute, 305-348 Daejeon, Korea}

\begin{abstract}
We present the discovery of the first Neptune analog exoplanet
or super-Earth with Neptune-like orbit, MOA-2013-BLG-605Lb. 
This planet has a mass similar to that of Neptune or a super-Earth 
and it orbits at $9\sim 14$ times the expected position of the snow-line,  $a_{\rm snow}$,
which is similar to 
Neptune's separation of $ 11\,a_{\rm snow}$ from the Sun. 
The planet/host-star mass ratio is $q=(3.6\pm0.7)\times 10^{-4}$ 
and the projected separation normalized by the Einstein radius is $s=2.39\pm0.05$. 
There are three degenerate physical solutions and two of these
are due to a new type of degeneracy in the 
microlensing parallax parameters, which we designate ``the wide degeneracy''. 
The three models have (i)
a Neptune-mass planet with a mass of $M_{\rm p}=21_{-7}^{+6} M_\earth$
orbiting a low-mass M-dwarf with a mass of $M_{\rm h}=0.19_{-0.06}^{+0.05} M_\sun$, (ii)
a mini-Neptune with $M_{\rm p}= 7.9_{-1.2}^{+1.8} M_\earth$ 
orbiting a brown dwarf host with $M_{\rm h}=0.068_{-0.011}^{+0.019} M_\sun$
and (iii) a super-Earth with $M_{\rm p}= 3.2_{-0.3}^{+0.5}  M_\earth$ 
orbiting a  low-mass brown dwarf host with $M_{\rm h}=0.025_{-0.004}^{+0.005} M_\sun$
which is slightly favored. 
The 3-D planet-host separations are 4.6$_{-1.2}^{+4.7}$ AU, 2.1$_{-0.2}^{+1.0}$ AU and 0.94$_{-0.02}^{+0.67}$ AU,
which are $8.9_{-1.4}^{+10.5}$,  $12_{-1}^{+7}$ or $14_{-1}^{+11}$  times larger than $a_{\rm snow}$ for these models, respectively.
The Keck AO observation confirm that the lens is faint.
This discovery suggests that low-mass planets with Neptune-like orbit are common. 
So processes similar to the one that formed Neptune
in our own Solar System or cold super-Earth may be common in other solar systems.
\end{abstract}


\section{Introduction}
The formation of  the ice giants Uranus and Neptune is not well understood.
In the favored core accretion theory, the gas
giant planets like Jupiter and Saturn are believed to form through
the accumulation of small icy planetesimals 
into solid cores of about 5-$15\,\mearth$ in the region beyond the 
snow-line at $a_{\rm snow}\approx 2.7(M_h/M_\sun)$  \citep{ida04,lau04,ken06},
where the protoplanetary disk is cold enough for ices (especially water-ice) to condense.
However, such a scenario is unable to form smaller ice giants like Uranus and Neptune 
at their current orbital positions,
due to the low density of planetesimals and slow evolution in these
orbits \citep{Pollack1996}.
One idea is that Uranus and Neptune formed in the 
Jupiter-Saturn region between $\sim$ 5 and $\sim$17 AU, 
then migrated outwards to the current position
 \citep{Fernandez1984,Thommes1999,Helled2014}. 

The formation of super-Earth exoplanets with Neptune-like orbit are even less understood. 
This is partly because we do not even know if they exist contrary to the case of Neptune. 
They are also not expected by the standard core accretion theory due to the same reason 
as that of ice giants mentioned above \citep{ida04}. Their formation may be similar to or 
related to the formation of Neptune-like ice giants.

The distribution of such cold ice-giant planets and super-Earth in other solar systems is important
for understanding the formation of our own cold ice giants. Also, in our
own Solar System, the distribution of Kuiper Belt objects (KBOs) is dominated by 
gravitational interactions with Neptune. Since KBOs hold
large amounts of water and other volatiles needed for life, it could be that 
exo-neptunes play an important role in the development of life in some exoplanetary
systems, whether or not they play this role in our own Solar System.

In the 20 years since the first exoplanet discovery \citep{may95}, there
have been repeated discoveries of planets that are quite different from
those in our own Solar System.
However, the detection of planets similar to those in our own Solar System has
been more difficult. Only Jupiter analogs have been detected orbiting solar type
stars \citep{Wittenmyer2014}, while 
Jupiter/Saturn \citep{gaudi-ogle109,bennett-ogle109} and Venus/Earth
analogs \citep{Burke2014,Quintana2014} have been found orbiting low-mass stars. Very
cold, low-mass planets have yet to be explored (see the distribution of 
known exoplanets as of 2015 Oct. 6 \footnote[1]{http://exoplanet.eu}  in Figure \ref{fig:a_Mp}).
Cold ice-giants like Uranus and Neptune are very difficult to detect 
with the radial velocity and transit methods owing to their long orbital periods
(80-160 years), low orbital velocities and low transit probabilities.
It is even more difficult to detect a super-Earth in such wide orbit.
The direct imaging method can detect wide-orbit planets if they are 
self-luminous, but otherwise, they will be far too faint to detect, especially
if they are as small as Neptune and super-Earth.

Recently, low-mass stars (i.e.\ M-dwarfs) have attracted more interest in exoplanet search
programs because of their high detectability of habitable or cold low-mass planets. 


\emph{Kepler}'s 150,000 targets contain about 3000 red dwarfs and more than 
a hundred planetary systems have been found orbiting these stars \citep{Morton2013}. 
These results show that smaller planets are more common than
larger planets around M-dwarfs, and planets 
with radii of $\sim$1.25 $R_\earth$ are the most common planets in these systems.
\cite{Dressing2013} estimated an occurrence rate of $\sim0.5$ habitable zone 
Earth size planets per M-dwarf, and \cite{Quintana2014} found an Earth-radius 
habitable planet around a $\sim$0.5 $M_\sun$ M-dwarf.
That smaller planets are more common than larger planets around M-dwarfs 
may be related to the fact that only small mass proto-planetary disks have been found
around such low mass stars \citep{Kennedy2008}.
The TRENDS high-contrast imaging survey, in combination with radial velocity measurements,
indicates that $6.5\% \pm 3.0\%$ of M-dwarf stars host one or more massive companions 
with $1 < m/M_{\rm J} < 13$ and $0 < a < 20$ AU,  however this survey is not sensitive to cold ice planets \citep{Montet2014,Clanton2014}.

The gravitational microlensing method is also sensitive to planets around M-dwarfs
and even brown dwarfs because it does not rely on the light from 
the host stars. Microlensing relies upon random alignments between background
source stars and foreground lens star+planet systems, and more massive
lens stars are only favored by the factor $\sqrt{M}$ while smaller masses have
shorter timescales which can also bias against detection. So M-dwarf lens stars dominate 
microlensing events.
Contrary to the other methods, microlensing is sensitive to low-mass planets 
down to an Earth-mass \citep{bennett96} orbiting beyond the snow-line, 
as shown in Fig. \ref{fig:a_Mp}. 
Microlensing is therefore complementary to the other planet detection techniques. Statistical analyses of
microlensing samples indicate that the planet abundance beyond the snow line
is about a factor $\sim 7$ larger than the abundance of close-in planets.
Neptune mass planets are more abundant
than gas giants around M-dwarfs, and one or more planets per star in total
are predicted just beyond the snow-line \citep{sumi2010,Gould2010,cassan2012}.

 In about half of the planetary systems found by microlensing, 
 the mass of the host and planets and their projected separation have 
 been measured by microlensing parallax in combination with the
 finite source effect \citep{bennett08,gaudi-ogle109,Muraki2011,Kains2013,Tsapras2014,Udalski2015} 
 and/or direct detection of the lens flux by 
 high resolution imaging by adaptive optics (AO) 
 \citep{bennett-ogle109,kubas2012,Batista2014,Batista2015} or the \emph{Hubble Space Telescope} 
 ({\it HST}) \citep{Bennett2006,bennett15,Dong2009-MOA400}.
 The probability distribution of physical mass and separations of other events have been estimated 
 using a Bayesian analysis assuming a Galactic model. 
Among the planetary systems with mass measurements,  two of them have
very low mass hosts, less than $0.2 M_\sun$ 
and each system has a planetary mass ratio $q<0.01$. These two systems are 
MOA-2007-BLG-192L ($M_{\rm h}= 0.084_{ -0.012}^{+0.015}M_\sun$, 
$M_{\rm p}=3.2_{-1.8}^{+5.2} M_\earth$) \citep{bennett08,kubas2012} and
MOA-2010-BLG-328L
($M_{\rm h}= 0.11\pm 0.01M_\sun,
M_{\rm p}= 9.2 \pm2.2M_\earth$)
\citep{furusawa2013}. 
Neptune analog planets and super-Earth with Neptune-like orbit are still difficult to detect even by microlensing.

Recently, \cite{Poleski2014} found a planet in a Uranus-like orbit with 
mass of $\sim$4 $M_{\rm Uranus}$ at $\sim$18 AU around $\sim$0.7 $M_\sun$
star. This is $\sim 9$ times the snow-line of the host. 
While their mass estimates are based on a Bayesian analysis and have large 
uncertainties, their detection demonstrated the ability to detect planets in these orbits with microlensing.

In this paper, we present the detection and the mass measurement of the first Neptune analog
MOA-2013-BLG-605Lb via microlensing. We detected the microlensing parallax effect which 
yield the mass measurement of the lens system in combination with the finite source effect.

Microlensing parallax can be measured when one observes an event simultaneously from two different locations,
either with a telescope on Earth and a space telescope, \citep{Refsdal1966,Udalski2015} or 
with two ground-based telescopes, referred to as terrestrial parallax \citep{Gould2009}.
It is known that there is a four-fold degeneracy in these parallax measurements, \citep{Refsdal1966,Gould1994}.
Two elements of this four-fold degeneracy correspond to two different magnitudes of the measured parallax.
 As a result, the physical parameters of the lens differs between these two degenerate solutions. 
 The other two degenerate solutions in the four-fold degeneracy just arise from a symmetry in the lensing geometry. 
 The physical parameters of the lens are the same between these two degenerate solutions,
except the projected velocities which can be used to distinguish
among solutions \citep{Novati2015}.
Most commonly, parallax measurements have been made by observing an event from an accelerated observatory; 
specifically from ground-based observations of an event which is long enough for Earth to move 
significantly in its orbit around the Sun. This is referred to as orbital parallax \citep{gou92}.
There is also an analogous four-fold discrete degeneracy for orbital parallax, termed the
``jerk parallax'' degeneracy and their mirror solutions \citep{Gould2004,Park2004}.
For the binary lens case, there is an approximate degeneracy in the parallax parameters, known as the ``ecliptic degeneracy''
\citep{Skowron2011}. 
In this work on event MOA-2013-BLG-605, we report a new type of degeneracy in parallax model solutions, which 
is specific to widely separated binary lenses. The details of this new degeneracy are presented in section \S\,\ref{sec:model}.

We describe the observations of, and photometric data for, event MOA-2013-BLG-605  in sections \S\,\ref{sec:observation}  and \S\,\ref{sec:data}.
The light curve modeling is described in section \S\,\ref{sec:model}.
In section \S\,\ref{sec:property} and \S\,\ref{sec:KeckAO} we present the physical parameters of the lens system
and constraints by the Keck AO observation.
We discuss, in section \S\,\ref{sec:future}, 
the manner in which we might measure the lens mass in the future
and we present an overall discussion and our conclusions in section \S\,\ref{sec:discussionAndSummary}.

\section{Observation}
\label{sec:observation}

The Microlensing Observations in Astrophysics (MOA; \citealt{bon01,sumi03})
collaboration carries out a microlensing survey toward the Galactic bulge 
from the Mt. John University Observatory in New Zealand. The MOA-II survey 
\citep{sumi2011} 
is a very high cadence photometric survey of the Galactic bulge 
with the 1.8\,m MOA-II telescope equipped with a 2.2 deg$^2$
field-of-view (FOV) CCD camera. The 2013 MOA-II observing strategy called for the 
6 fields ($\sim13\,{\rm deg}^2$) with the highest lensing rate to be observed with a 15 minute 
cadence, while the next 6 best  fields were observed with a 47 minute cadence, and
8 additional fields were observed with a 95 minutes cadence. Most MOA-II observations 
use the custom MOA-red wide band filter, which corresponds to the sum of the standard 
Cousins $R$ and $I$-bands. MOA-II issues $\sim 600$ alerts of microlensing 
events in real time each year.\footnote[2]{https://it019909.massey.ac.nz/moa/}

The Optical Gravitational Lensing Experiment (OGLE; \citealt{UdalskiSzymanski2015}) also
conducts a microlensing survey toward the Galactic bulge with the 1.3 m Warsaw telescope
at the Las Campanas Observatory in Chile.
The fourth phase of OGLE, OGLE-IV started its high cadence survey observations in 
2010 with a 1.4 deg$^2$ FOV mosaic CCD camera. OGLE observes bulge fields
with cadences ranging from one observation every 20 minutes for 3 central
fields to less than one observation every night for the outer bulge fields.
Most observations are taken in the standard Kron-Cousin $I$-band with 
occasional observations in the Johnson $V$-band.
OGLE-IV issues $\sim 2000$ microlensing event alerts in real time 
each year.\footnote[3]{http://ogle.astrouw.edu.pl/ogle4/ews/ews.html}

\begin{figure}
\begin{center}
\includegraphics[angle=-90,scale=0.7,keepaspectratio]{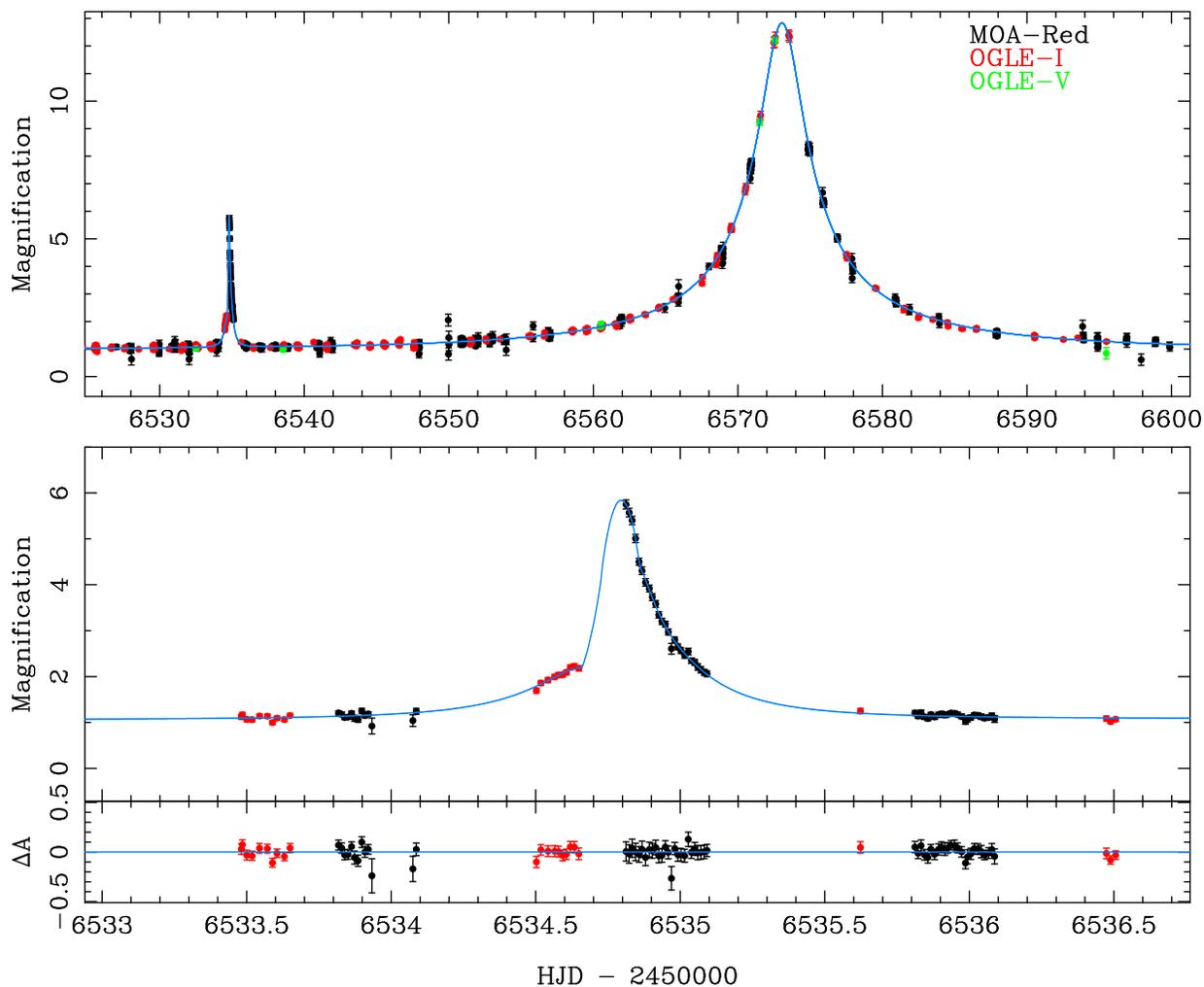} 
\caption{
  \label{fig:lcurve}
Light curve of MOA-2013-BLG-605. Black, red and green points indicate MOA-Red, OGLE-I and 
OGLE-V band data, respectively. Blue lines represent
the parallax model with Keplerian orbital motion, 
a Keplerian prior and the kinematic constraint P$_{++}$K$_{\rm pk}$ 
which is almost identical to all models with parallax and orbital motion.
Middle and Bottom panels show the detail of the planetary 
signal and its residual from the best model.
}
\end{center}
\end{figure}

The microlensing event MOA-2013-BLG-605 was discovered at
($\alpha$, $\delta$)(2000)
= (17:58:42.85, -29:23:53.66) [($l,b$) = (1.0583$^\circ$, -2.695$^\circ$)],
in MOA field gb9, which is monitored every 15\,min, and it was
announced by the MOA Alert System on 2013 Aug 30
(HJD$' \equiv$HJD$-2450000\sim 6535$).  
Figure \ref{fig:lcurve} shows the light curve.
At the time of its discovery, MOA recognized this event as a possible
free-floating planet candidate  \citep{sumi2011} as the best fit single lens light curve 
had an Einstein radius crossing time of $t_{\rm E} = 0.73 \pm 0.10\,$days 
(See Figure \ref{fig:lcurve}). Nearly four weeks later, the
OGLE Early Warning System (EWS)  \citep{Udalski2003}  detected
this event being magnified again with longer timescale due to the lensing
effect of the host star. The OGLE EWS system announced this event
as OGLE-2013-BLG-1835 on 2013 Sep 25 (HJD$' \sim6560$),
as shown in the top panel of Figure \ref{fig:lcurve}.
The initial short magnification by the planet at HJD$'\sim 6535$ 
was confined by the OGLE survey data. Actually, it should have
triggered the OGLE discovery alert but due to unfortunate deeply hidden
bug in the EWS software this did not happen. 
The later magnification by the host 
was observed by MOA, as well as OGLE.

Follow-up observations of the stellar part of the light curve in the $V$, $I$ and 
$H$-bands were obtained by the $\mu$FUN collaboration using the
SMARTS-CTIO 1.3\,m  telescope.
These data were taken mainly to extract the source color.
We use the average of 
these CTIO and OGLE $V-I$ color measurements.
CTIO $H$-band measurements are used to drive $H$-band source magnitude,
which is very important for comparison to the AO observations (see Section \ref{sec:KeckAO}).

\section{Data Reduction}
\label{sec:data}

The MOA images were reduced with MOA's implementation \citep{bon01} of 
the difference image analysis (DIA) method \citep{tom96,ala98,ala00}. 
In the MOA photometry, we found that there were systematic errors 
that correlate with the seeing and airmass, as well as the motion,
due to differential refraction, of a nearby, possibly unresolved star.
There is also a potential systematic error due to
the relative proper motion of a nearby star or stars, which we model as
linear function in time. We ran a detrending code to measure these
 effects in the 2011, 2012 and 2014 data, and we removed these
trends with additive corrections to the full 2011-2014 data set. (The MOA
data from 2006-2010 indicate no significant photometric variations, but
they are not included in the light curve analysis.)
This detrending procedure improved the fit $\chi^2$ by 
$\Delta\chi^2=  0.073$ per data point in the baseline, so it has reduced
the systematic photometry errors significantly. This investigation of the systematics
is necessary to have confidence in the modeling of the
light curve with high order effects in the following section.

The OGLE data were reduced with the OGLE DIA \citep{wozniak2000}
 photometry pipeline \citep{UdalskiSzymanski2015}. 
In this event, the center of the magnified source star is slightly shifted from the 
center of the apparent star identified in the reference image, due to blending 
with one or more unresolved stars. So the OGLE data have been re-reduced 
with a centroid based difference images, just as the MOA pipeline does \citep{bon01}.

The number of data points used for the light curve modeling are 9675, 5514 
and 64 for MOA-Red, OGLE-$I$ and OGLE-$V$ passbands, respectively. 
The photometric errorbars provided by the photometry codes give approximate 
estimates of the absolute photometric uncertainty of each measurement, and we 
regard them as an accurate representation of the relative uncertainty for each
measurement. This is adequate for determining the best light curve model, but in
order to determine the uncertainties on the model parameters, it is important to
have more accurate error bars. We accomplish this with the method presented
in \citet{yee2012}. We rescale the errors using the formula,
$\sigma_i^{\prime}=k\sqrt{\sigma_i^2 +e_{\rm min}^2}$,
where $\sigma_i$ and $\sigma_i^{\prime}$ are original and renormalized 
errorbars in magnitudes. The parameters $k$ and $e_{\rm min}$ are selected so 
that the cumulative $\chi^2$ distribution sorted by the magnification of the 
best model is a straight line of slope 1 and $\chi^2/$dof$\sim 1$. This procedure
yields $k = 1.092313$ and $e_{\rm min} = 0.012662$ for MOA-Red, 
 $k = 1.387059$ and $e_{\rm min} = 0.010938$ for OGLE-$I$ and
 $k = 1.571492$ and $e_{\rm min} = 0.0$ for OGLE-$V$.
Note that the changes of the final best fit model due to this error renormalization
are negligible. 

CTIO data were reduced by DoPHOT \citep{sch93},
the point spread function (PSF)-fitting routine.
The number of data points in the CTIO-I, V and H pass-bands is 15, 15 and 149, respectively.
Their error bars are not rescaled, i.e., 
 $k = 1.0$ and $e_{\rm min} = 0.0$. 
 These CTIO data are not used for light curve modeling,
 but used for obtaining the source color in $(V-I)$ and $(I-H)$
 in a model-independent way from the linear regression of these
 light curves
by following \cite{Dong2009} and  \cite{Novati2015photometory}.
  
Details of the datasets are summarized in Table \ref{tbl:dataset}.

\section{Light curve modeling}
\label{sec:model}

We search for the best fit models of the standard (static), the parallax, the parallax with 
the linear orbital motion of the planet, the keplerian orbital motion, with the keplerian prior, 
the Galactic kinematic constraint and the Galactic density prior using Markov Chain Monte Carlos (MCMC) \citep{verde03}. 
The best fit models are shown in Table~\ref{tbl:param}-\ref{tbl:param26} and
their physical parameters are in Tables \ref{tbl:physical}-\ref{tbl:physical26}, 
respectively (see Section \ref{sec:property}).

 \subsection{Standard (static) model}
 \label{sec:static}
 
In a point-source point-lens (PSPL) microlensing model, there are three  parameters, 
the time of peak magnification $t_0$, the Einstein radius crossing time $t_{\rm E}$, 
and the minimum impact parameter $u_{0}$. The standard binary lens model has 
four more parameters, the planet-host star mass ratio $q$, the projected separation 
normalized by Einstein Radius $s$, the angle of the source trajectory relative 
to the binary lens axis $\alpha$, and the ratio of the angular source radius to the angler 
Einstein radius $\rho=\theta_*/\theta_{\rm E}$. $\rho$ can only be
measured for events that show finite source effects.
The measurement of $\rho$ is important because
it allows us to determine the angular Einstein radius 
$\theta_{\rm E} = \theta_{\rm *}/\rho$ since the angular source radius, 
$\theta_{\rm *}$, can be estimated from its color and extinction-corrected apparent 
magnitude \citep{kervella2004}.

We use linear limb-darkening models for the source star
using the coefficients, $u=0.5863$, $0.7585$ and $0.6327$ for the $I$, $V$ and 
MOA-Red bands, respectively \citep{Claret2000}. The MOA-Red value is the
mean of the $R$ and $I$-band values. These values were selected
from \cite{Claret2000} for a
K2 type source star with $T=5000$ K, log$g=4.0$ and $\log[M/H]=0$, based
on the extinction corrected, best fit source $V-I$ color and brightness 
(see Section \ref{sec:property}).

Initially, the global grid search of the best fit model was conducted with 9,680 fixed grid points
across a wide range of three parameters, $-4.0  <\log q <0.4$, $-0.5  <\log s <0.6$ and $0< \alpha < 2\pi$,
with all other parameters being free. 
Then the most likely models were refined, allowing all parameters to vary. Using this robust
search methodology, we avoid missing any local minimum solutions across the wide range of parameter space.
We found that only the model with a wide separation ($s>1$) reproduces the observed light curve data.
The model corresponds to the source crossing a planetary caustic. Planetary caustics can 
form far from the primary and any source star that crosses or passes close to such a 
distant a planetary caustic will impose an signal far from the main microlensing peak.
Furthermore, the shape of  planetary caustics differ significantly between wide ($s>1$) 
and close ($s < 1$) configurations in contrast to the close/wide degeneracy for an event 
crossing a central caustic near the primary.
The best fit standard model parameters are shown in Table \ref{tbl:param}.
The mass ratio of $q\sim 3\times 10^{-4}$ and separation of $s\sim2.3$
indicate that the companion is a relatively low mass planet at wide separation.

The single lens model with a binary source was ruled out as follows.
We extracted 88 data points around the planetary anomaly within $6533.0<$HJD$'< 6536.6$
after subtracting the flux contribution from the best fit single lens model of the primary peak 
which is fitted without data around the anomaly.
We fitted this extracted light curve by the single lens model with a finite source effect.
The best fit $\chi^2$ is $\sim$31 larger than the $\chi^2$ contributions from 
the same data points by the planetary models with parallax and orbital motion K$_{\rm pk}$ (see \S \ref{sec:kinematic}). 
Furthermore, the best fit event timescale 
of $t_{\rm E}=2.3$\,days is much smaller than $t_{\rm E}\sim 20$\,days of the main peak,
while they should be same if a single lens caused both two magnifications.

 \subsection{Parallax model with a New Type of Degeneracy}
 \label{sec:parallax}

There are higher order effects that require additional parameters. 
The orbital
motion of the Earth can cause the apparent lens-source relative motion to
deviate from a constant velocity.
 This effect is known as the microlensing
parallax effect \citep{gou92,alc95,smi02}, and it can be described by the
microlensing parallax vector \mbox{\boldmath $\pi$}$_{\rm E}=(\pi_{\rm E,N}, \pi_{\rm E,E})$.
The direction of \mbox{\boldmath $\pi$}$_{\rm E}$ is the direction of the lens-source 
relative motion projected on the sky (geocentric proper motion at a fixed time), and the amplitude of the
microlensing parallax vector, $\pi_{\rm E}={\rm AU}/\tilde{r}_{\rm E}$, is the
inverse of the Einstein radius, projected to the observer plane. Because the
Galactic bulge is close to the ecliptic plane, there is an approximate degeneracy
in the parallax parameters, known as the ``ecliptic degeneracy\rlap," where models
with similar parameters but with $(u_0,\alpha,\pi_{\rm E,N}) =-(u_0,\alpha,\pi_{\rm E,N}) $
produce nearly indistinguishable light curves. This corresponds to a reflection of the lens 
plane with respect to the geometry of Earth's orbit,
\citep{Smith2003,Skowron2011}.

\begin{figure}
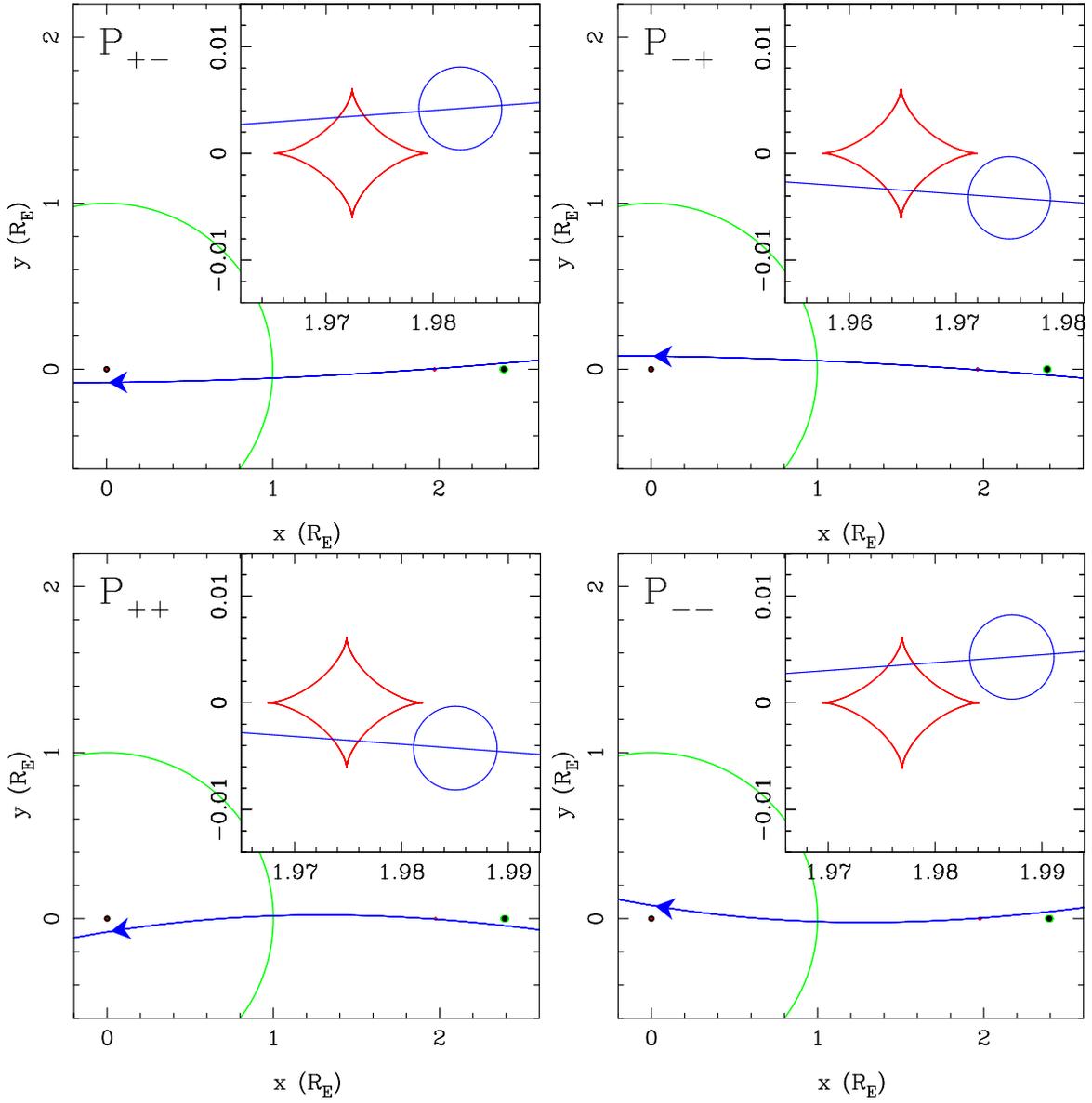

\begin{center}
\includegraphics[angle=-90,scale=0.4,keepaspectratio]{Figure3a.eps} 
\includegraphics[angle=-90,scale=0.4,keepaspectratio]{Figure3b.eps} 
\includegraphics[angle=-90,scale=0.4,keepaspectratio]{Figure3c.eps} 
\includegraphics[angle=-90,scale=0.4,keepaspectratio]{Figure3d.eps} 
\caption{
  \label{fig:caustic}
Caustics  (Red lines) of MOA-2013-BLG-605 of the best fit 
parallax models,
P$_{+-}$, P$_{-+}$, P$_{++}$ and P$_{--}$
in top-left, top-right, bottom-left and bottom-right panel, respectively.
The figure of each model with a orbital motion and various priors are similar. 
Insets show a close-up view around the planetary caustic.
Blue circles indicate the best fit source star radius and position at HJD= 6534.6, just before crossing the planetary caustic.
Blue lines with arrows represent source star trajectories. 
The left and right black filled circles at $y=0$ indicate the positions of primary and planet, respectively.
The green lines show critical curves.
}
\end{center}
\end{figure}

We found the four degenerate parallax models as shown in Table \ref{tbl:param}. 
The light curves of these four models are almost identical to the one shown in Figure \ref{fig:lcurve}.
The caustics, critical curves and source trajectory of these models are shown in Figure \ref{fig:caustic}.
The ``P" scripts indicate models with microlensing parallax. 
The "$+$" and "$-$" subscripts refer to two different 2-fold degeneracies
in the parallax models. The first ``$\pm$" subscript refers to the sign of the $u_0$ parameter,
and refers to the  ``ecliptic degeneracy" mentioned above, but the second ``$\pm$" 
subscript refers to a new parallax degeneracy, the wide degeneracy, that is particular to events
like this, with a wide separation planet detected through a crossing of the planetary caustic. 
The light curve measurements indicate the angle, $\alpha(t_{\rm pcc})$,
between the source trajectory and lens axis at the time of the planetary caustic
crossing, $t_{\rm pcc}$. Due to the reflection symmetry of the lens system, the
light curve constrains $\alpha(t_{\rm pcc})$  up to a reflection symmetry, as
shown in Figure~\ref{fig:caustic}. If there were no microlensing parallax,
we could use $\alpha(t_{\rm pcc})$ to predict the closest approach of the source
to the center-of-mass, $u_0$, and therefore the peak magnification of the
the stellar part of the microlensing light curve. But, when the microlensing parallax
effect is included, the angle $\alpha$ can vary in time, so that 
$\alpha(t_{\rm pcc}) \neq \alpha(t_0)$. For a wide-separation planetary event,
like MOA-2013-BLG-605, the light curve basically constrains 
the microlensing parallax through the three parameters, $\alpha(t_{\rm pcc})$, $u_0$,
and $t_0$, which is essentially the time of the stellar peak magnification (cf. \citealt{An2001}).
As a result, the configurations shown in the upper and lower panels of
Figure~\ref{fig:caustic} yield nearly identical light curves as shown in Figure \ref{fig:lcurve}, 
even though the source passes through in between the two masses in the upper panels
and below or above the masses in the bottom two panels. The lower panels
imply a larger curvature of the source trajectory, and therefore, a larger
microlensing parallax signal. 
(Note that the model parameter 
$\alpha_0$ and $s_0$ given in Table~\ref{tbl:param}-\ref{tbl:param26} are
the $\alpha$ and $s$ values at a fixed 
time $t_{\rm fix} = 6573.045$, following the convention of Geocentric
microlensing parallax parameters.)

\begin{figure}
\begin{center}
\includegraphics[angle=0,scale=0.5,keepaspectratio]{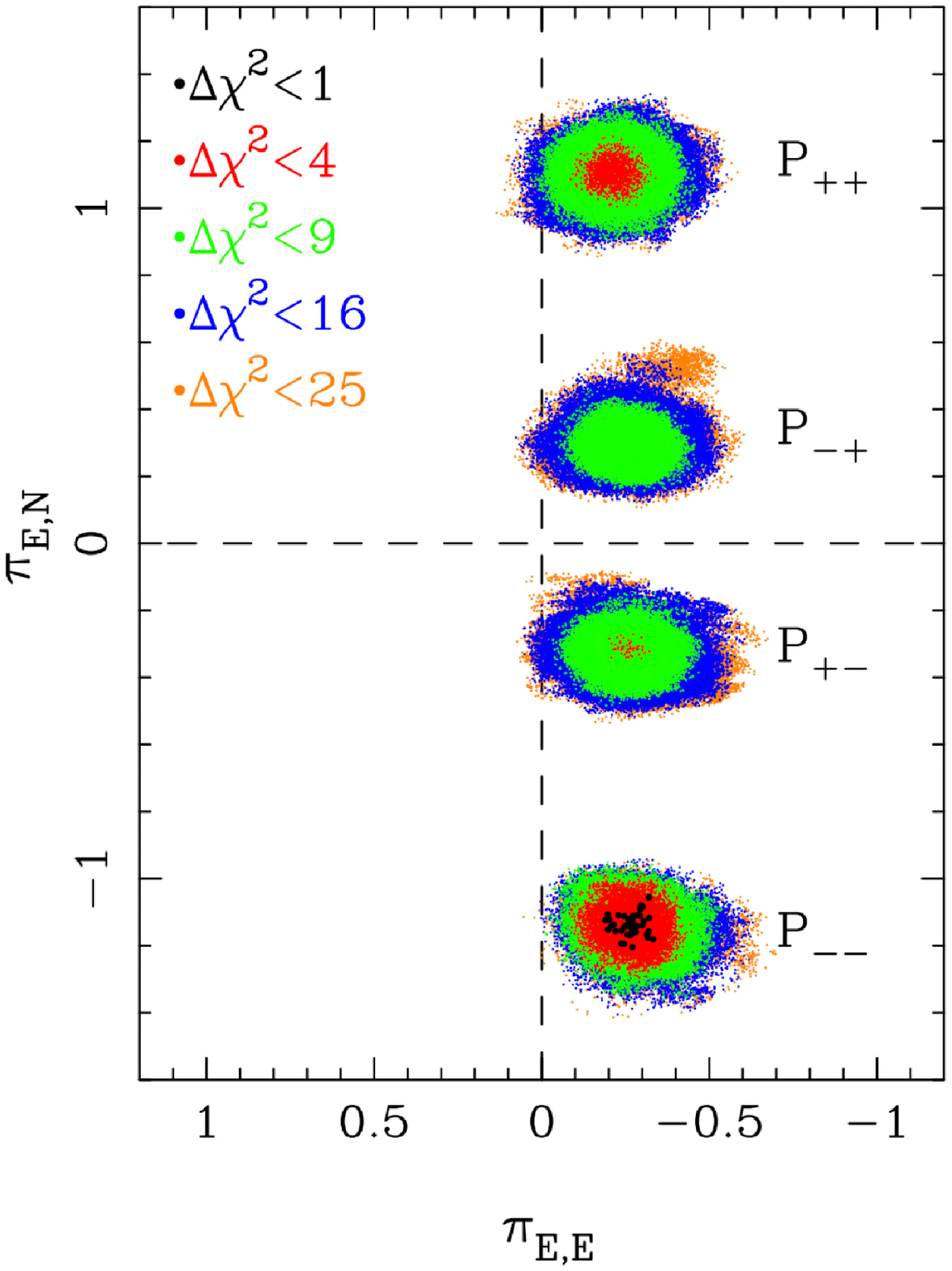} 
\caption{
  \label{fig:chains}
$\Delta\chi^2$ distribution of the parallax parameters from the best fit parallax-only models,
without orbital motion, 
P$_{++}$,  P$_{-+}$, P$_{+-}$ and P$_{--}$,  from top to bottom. 
Black, red, green, blue and orange dots indicate chains with $\Delta \chi ^2 <1$, $4$, $9$, $16$ and $25$,
respectively. The black nots are larger than the others for clarity.
}
\end{center}
\end{figure}

Fig. \ref{fig:chains} shows the $\Delta\chi^2$ distribution of the parallax parameters from the best fit MCMC models.
The best fit values are compared to that of other models in Figure \ref{fig:piENE}.
 In late September, the Earth's acceleration is in the East-West (E-W) direction, so for
a typical event, we would expect a better constraint on  parallax in the E-W direction, i.e, 
a smaller error for $\pi_{\rm E,E}$. 
However, in this case, the planetary signal plays a big role in the parallax signal. 
The angle and timing of caustic entry for a given $u_0$ value -- which is constrained by 
the main peak corresponding to the host star -- constrain the parallax parameters.
 
\begin{figure}
\begin{center}
\includegraphics[angle=-90,scale=0.7,keepaspectratio]{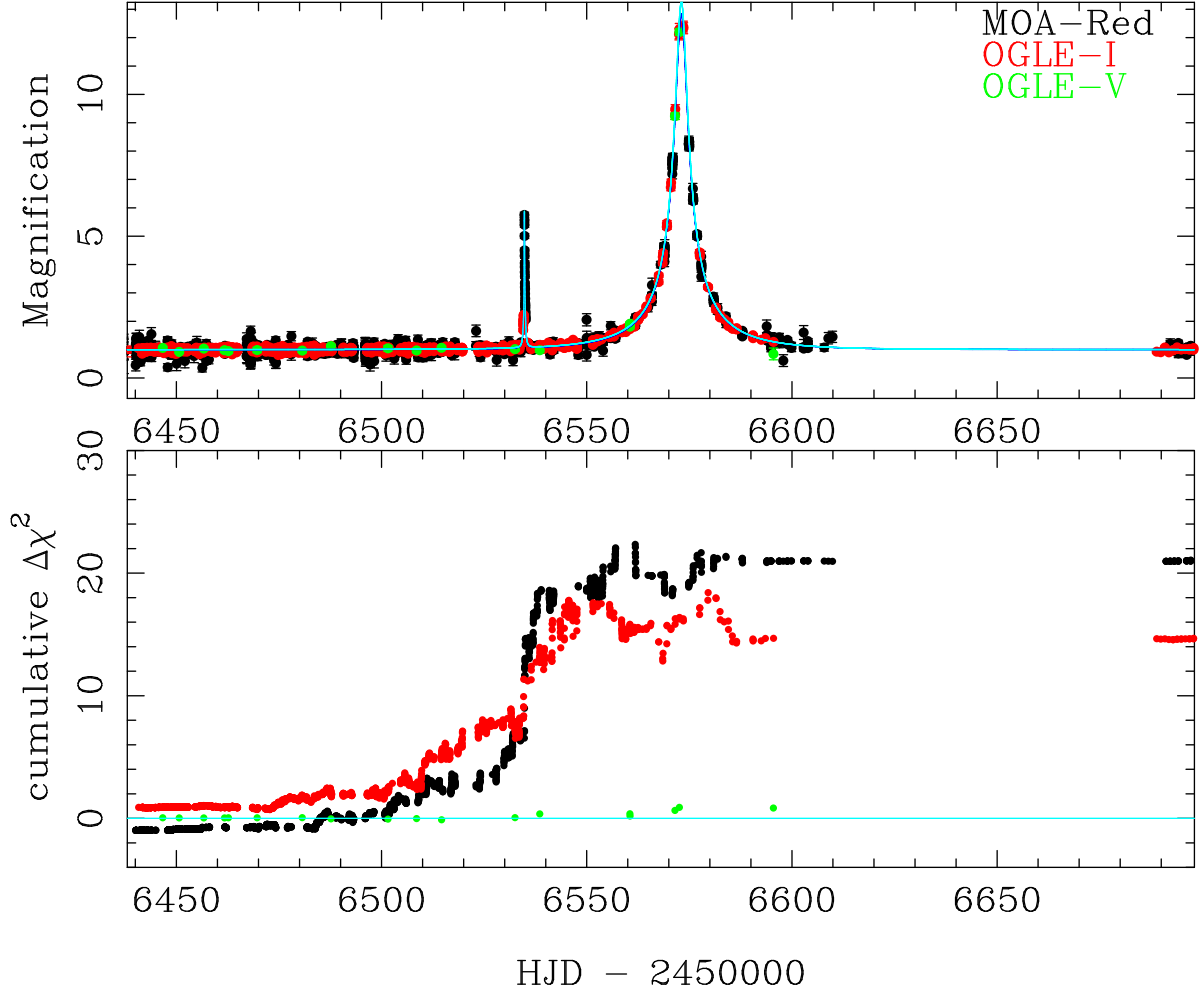}
\caption{
  \label{fig:dcumchi2}
The difference in the cumulative $\chi^2$ values between the
standard and the parallax models, $\chi^2_{\rm standard} - \chi^2_{\rm parallax}$, 
is shown as a function of time in the bottom panel.
This is the case of the parallax model with an orbital motion, keplerian prior and  kinematic prior, 
P$_{++}$K$_{\rm pk}$ (see section \S \ref{sec:kinematic}), but it is similar for all models with parallax.
The corresponding light curve is given in the top panel as a reference.
Blue and cyan solid lines represent the standard and P$_{++}$K$_{\rm pk}$ models, while 
the cyan line almost overlaps the blue one.
Black, red and green points represent MOA-Red, OGLE-$I$ and OGLE-$V$, respectively.
The plots are slightly shifted to match at the beginning of 2013 season for clarity, 
where the total  $\Delta\chi^2$ in the baseline outside of the figure is almost zero. 
We can see that most of the parallax signals come from around the planetary signal in both MOA-Red and OGLE-$I$
as expected. This consistency in two datasets support the reality of the parallax signal.
}
\end{center}
\end{figure}

Figure \ref{fig:dcumchi2} shows the difference in the cumulative $\chi^2$ values between 
the standard and the parallax models as a function of time. 
We can see that most of the parallax signals come from around the planetary signal 
in both MOA-Red and OGLE-$I$ as expected.

For all these models, $q$ and $s$ are similar to that of the standard model, so the
companion is a cold low mass planet. For all 4 degenerate solutions,
the model parameters of greatest interest are all very similar, except for the
microlensing parallax. The ecliptic degeneracy yields nearly identical physical
parameters, except that the direction of the lens-source relative motion is
different. Potentially, this angle can be measured with follow-up observations
\citep{bennett15,Batista2015}. In contrast to the ecliptic degeneracy, the wide
degeneracy implies different amplitudes, $\pi_{\rm E}$, of the microlensing
parallax vector, which implies different lens system masses, as discussed
in Section~\ref{sec:property} below. 

Thus, this wide degeneracy presents us
with two different classes of physical models, P$_{\pm\mp}$ and P$_{\pm\pm}$,  
where the P$_{\pm\pm}$
models have larger $\pi_{\rm E}$ implying smaller lens system masses and
distance (see Section \ref{sec:property} and Table \ref{tbl:physical}).

These four parallax models are preferred  over the standard $\pi_{\rm E} = 0$ model  
in both the MOA-Red and OGLE-$I$ bands by 
 ($\Delta \chi^2_{\rm MOA},  \Delta \chi^2_{\rm OGLE})=
(-21.8,-11.5)$, $(-21.6, -11.1)$,  $(-21.9, -12.5)$  and $(-23.6, -12.8)$ 
for theP$_{+-}$, P$_{-+}$, P$_{++}$ and P$_{--}$  models, respectively. In total,
the $\chi^2$ differences range from $\Delta\chi^2 = -33.3$  to $-37.1$.
In Figure \ref{fig:dcumchi2}, one can also see that these $\chi^2$ improvements 
came from the same region of the light curves around the planetary anomaly 
in both datasets.
Microlensing parallax signals can sometimes be mimicked by the systematic
errors in the light curve photometry, but a consistent signal seen in both
the MOA and OGLE data implies that the signal is likely to be real.

 \subsection{Xallarap model}

The xallarap effect is a light curve distortion caused by the orbital motion of the 
source star \citep{Griest92,han97}, so it only occurs if the source star has a binary
companion \citep{derue99,alc01}.
Xallarap can be represented by five additional model parameters.
The xallarap vector \mbox{\boldmath $\xi$}$_{\rm E}=(\xi_{\rm E,N}, \xi_{\rm E,E})$ is similar to the parallax vector,  \mbox{\boldmath $\pi$}$_{\rm E}$, and
represents the direction of the lens-source relative motion. The amplitude of the xallarap vector, $\xi_{\rm E}=a_s/\hat{r}_{\rm E}$ is the semimajor axis of the source's orbit, $a_s$,
in units of the Einstein radius projected on the source plane,
$\hat{r}_{\rm E}= \theta_{\rm E}D_s$. The other xallarap parameters are the direction of the observer
relative to the source orbital axis, with vector components R.A.$_\xi$ and decl.$_\xi$, and the source binary orbital period, $T_\xi$.
For an elliptical orbit, two additional parameters are required, the orbital eccentricity,
$\epsilon$ and time of perihelion, $t_{\rm peri}$, which we did not consider here as their inclusion did not
improve the fit of the model to the data.

We found xallarap models giving only marginally better $\chi^2$ values compared to parallax models for  $T_\xi \ge160$ days and 
worse values of $\chi^2$ for shorter values of $T_\xi$.
This is not surprising as it is known that xallarap effects can mimic parallax effects \citep{Smith2003,Dong2009}.
Including xallarap yields a slight improvement of $\Delta \chi^2 \sim-5$ for  $160 \le T_\xi < 200$ days and
 $\Delta \chi^2 \sim-9$ at $T_\xi \ge200$ days. 
However, these models lead to a 
xallarap amplitude of $\xi_{\rm E} \ge 0.26$, which is larger than would be 
induced by a ``normal'' main-sequence companion. Here
$\xi_{\rm E}$ is expressed, making use of Kepler's third law, by
\begin{equation}
\xi_{\rm E}={a_s \over \hat{r}_{\rm E}}
    ={{1\rm AU}\over \hat{r}_{\rm E}} 
  \left(  {M_c\over M_\sun } \right)  \left( {M_\sun \over M_c  + M_s} {T_\xi \over 1 \rm yr} \right)^{2\over3}.
\label{eq:xiE}
\end{equation}
These models require a source companion of mass 
$M_{\rm c}> 6 M_\sun$ for $T_\xi \ge160$ days and 
$M_{\rm c}> 40 M_\sun$ for $T_\xi \ge 200$ days.
Such a heavy object would most likely be a stellar remnant or a black hole -- in either case, a rare object and thus an unlikely source companion. For this reason we reject the inclusion of the xallarap in our models.

 \subsection{Orbital motion model}
\label{sec:orbit}

 \subsubsection{Linear Orbital Motion}
\label{sec:LinearOrbitalMotion}

\begin{figure}
\begin{center}
\includegraphics[angle=-90,scale=0.6,keepaspectratio]{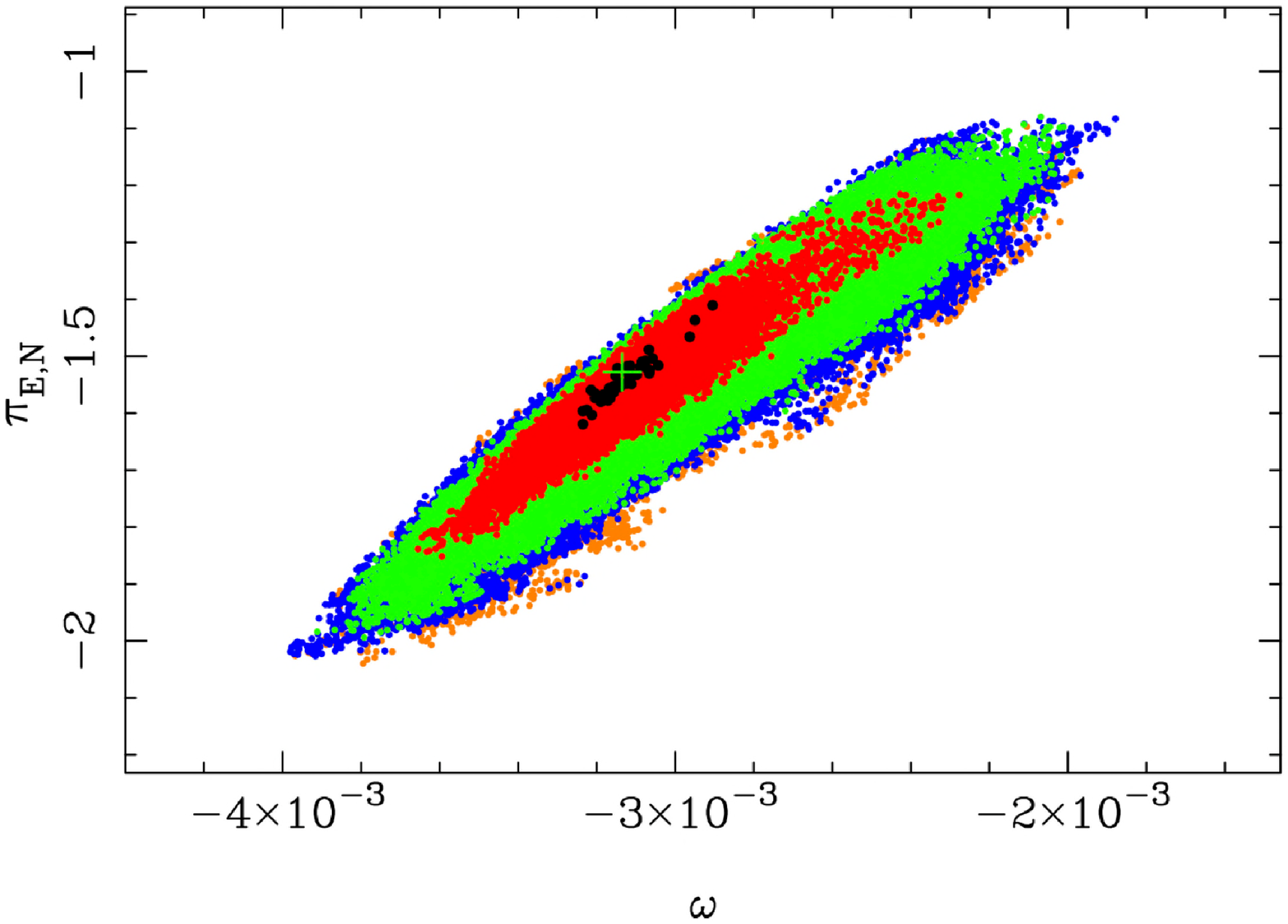} 
\caption{
  \label{fig:chains_degenerate}
$\Delta\chi^2$ distribution of the parallax parameter $\pi_{\rm E,N}$ and orbital motion perimeter $\omega$ 
from the best fit MCMC models of the P$_{+-}$L, the parallax with a linear orbital motion model.
Black, red, green, blue and orange dots indicate chains with $\Delta \chi ^2 <1$, $4$, $9$, $16$ and $25$,
respectively. Here $\omega$ is in rad\,day$^{-1}$ as actually used in the MCMC.
The black nots are larger than the others for clarity.
}
\end{center}
\end{figure}

The orbital motion of the planet around the host star causes a similar effect as parallax. 
To a first-order approximation, the orbital motion of the planet is described by two parameters,
the rate of change, $\omega=d\alpha/dt$ (radian yr$^{-1}$), of the binary axis angle $\alpha$,  
and the rate of change d$s$/${\rm d}t$ (yr$^{-1}$), of 
the projected lens star and planet separation $s$  \citep{Dong2009,Batista2011}, as follows,

\begin{equation}
s= s_0 + ds/dt (t-t_{\rm fix}),\, \alpha = \alpha_0 + \omega(t-t_{\rm fix}),
\label{eq:s_alpha}
\end{equation}
where, $s_0$ and $\alpha_0$  are instantaneous value of $s$ and $\alpha$ at the time   $t_{\rm fix}$.
We required the planet to be bound. That is, the ratio of the projected kinetic energy and potential energy,

\begin{equation}
\left( {\rm KE \over PE} \right)_\perp={(r_\perp / {\rm AU})^3 \over 8\pi^2 (M/M_\sun)} 
\left[  \left(  {1 \over s}{ds \over dt} \right)^2  +    \left(  {d\alpha \over dt} \right)^2      \right] {\rm yr}^2,
\label{eq:KEPE}
\end{equation}
which is less than the ratio of kinetic to potential energy (KE/PE) in three dimensions, 
was required to be less than unity in the MCMC calculations used to determine the model parameter distributions.
The four best linear orbital motion models (with scripts  "L") that correspond to each of four parallax models in  Table \ref{tbl:param}, are
shown in Table \ref{tbl:param1}. One finds that  $\pi_{\rm E}$ and its uncertainty significantly increased,
while the $\chi^2$ only slightly improved.
This is because of the well known degeneracy between one component of the parallax vector, 
$\pi_{\rm E,\perp}$, which is the perpendicular to the binary axis and close to  $\pi_{\rm E,N}$
in this case, and the lens orbital rotation on the sky, $\omega$. 
As an example,  $\Delta \chi^2$ distribution of $\pi_{\rm E,N}$ and $\omega$ for the model 
${\rm P_{+-}L}$ is shown in Figure \ref{fig:chains_degenerate}.

Note that there are two additional degenerate models ${\rm P_{\pm\pm}L'}$ which have smaller $s_0\sim1.97$  
and larger $ds/dt\sim-4.1$ yr$^{-1}$ compared to the other models. Here, $s$ of these models are
similar to others, $s\sim 2.4$, when the source crosses the planetary caustic.
However these models are disfavored with the full Keplerian orbit in the following analysis.

The physical parameters of the lens system of these models are shown in 
Table \ref{tbl:physical1} (see details in section \S \ref{sec:property}).
The host stars in these four models have a brown dwarf mass.
Note that ${\rm (KE / PE)_\perp}$ of these models given in Table \ref{tbl:physical1} are close to unity. 
The probability of having such high value is quite low as it requires very large eccentricity 
of $e\sim 1$, seeing the orbital plane face-on. 
If the parameters are not well constrained by the light curve, the density distribution of the 
MCMC chain depends on the prior probability of the fitting parameters in MCMC.  
Although the linear approximation of the lens motion is good enough in most of the cases, 
this parameterization inadvertently assumed the uniform prior on all microlensing fitting parameters,
which is not physically justified. We need to use a full keplerian orbit parameterization
to introduce physically justified priors.

 \subsubsection{Full Keplerian Orbit}
\label{sec:FullKeplerianOrbit}

To take the proper weighting on the orbital parameters, we adopt the full Keplerian 
parameterization by \cite{Skowron2011}. 
The advantage of the full Keplerian orbit is not only being more accurate and allowing only bound orbital solutions, 
it also enables us to introduce  physically justified priors on the orbital parameters.
In addition to the parameters defined above, we introduce the position and velocity along the line of sight, 
$s_{\rm z}$ in units of $r_{\rm E}$ 
and $ds_{\rm z}/dt$ in  yr$^{-1}$.
Then, the three dimensional position and velocity of the secondary relative to its host can be described by 
$(s_0, 0, s_{\rm z})$ and $s_0(\gamma_\parallel,\gamma_\perp,\gamma_{\rm z})=( ds/{\rm d}t , s_0\omega, ds_{\rm z}/dt )$.

We run MCMC fitting using the microlensing parameters with these six instantaneous Cartesian phase-space coordinates, 
in which we transform the ``microlensing'' parameters to ``Keplerian'' parameters, i.e., eccentricity ($e$), 
time of periapsis ($t_{\rm peri}$),
semi-major axis ($a$) and three Euler angles, longitude of the ascending node ($\Omega_{\rm node}$), 
inclination ($i$), and argument of periapsis ($\omega_{\rm peri}$). 
By following \cite{Skowron2011}, we assume flat priors on values of eccentricity, time of periapsis, log($a$), and $\omega_{\rm peri}$.  
Owing to the fact that orbital orientation is random in space, we multiply the prior by $|\sin i|$.
We must multiply the Jacobian of the parameter transformation function, 
$j_{\rm kep}=||\partial (e,a,t_{\rm peri}, \Omega_{\rm node},i, \omega_{\rm peri} )/ \partial (s_0, \alpha_0,s_{\rm z}, \gamma_\parallel,\gamma_\perp,\gamma_{\rm z})||$ (Eq. B6 in \citealt{Skowron2011}).
So we adopt the Keplerian orbit prior of $P_{\rm kep}=j_{\rm kep}|\sin i| a^{-1}$ and added the $\Delta\chi^2$ penalty of
$\Delta \chi^2_{\rm kep} = -2 \ln(P_{\rm kep})$.

\begin{figure}
\begin{center}
\includegraphics[angle=-90,scale=0.6,keepaspectratio]{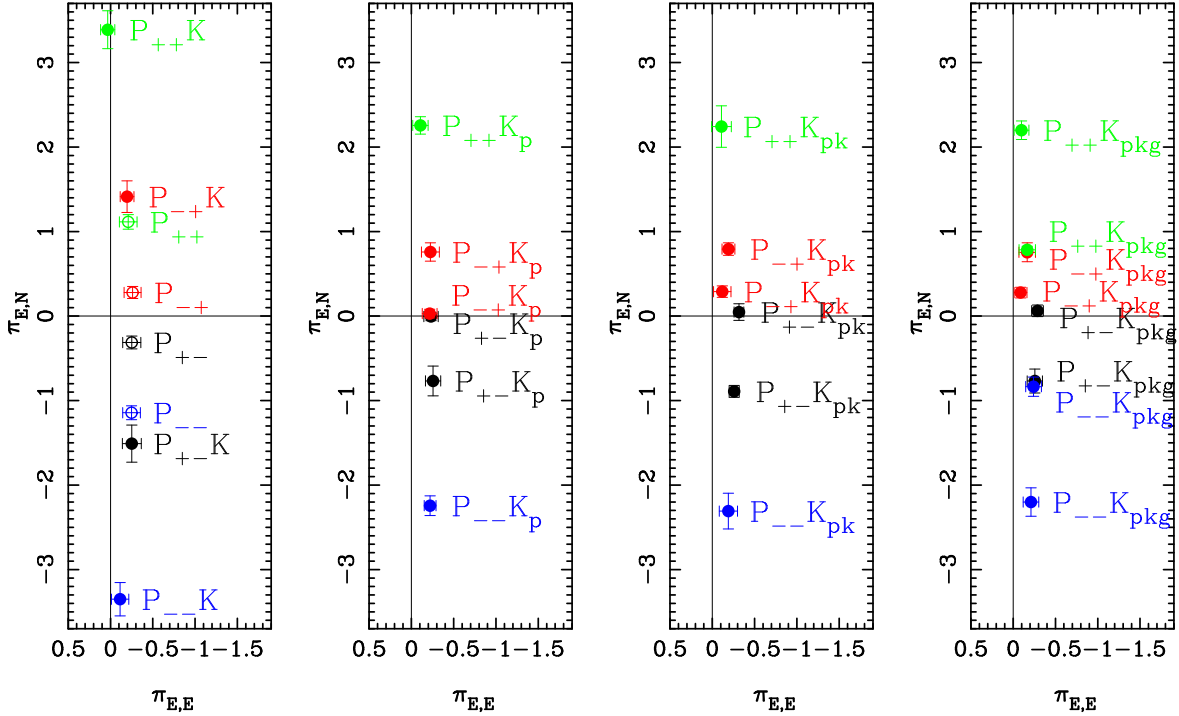}
\caption{
  \label{fig:piENE}
The best fit parallax parameters ($\pi_{\rm E,E}$, $\pi_{\rm E,N}$) of the parallax-only model (open circles in the left panel)
and the parallax with the orbital motion (filled circles) with the Keplerian orbit (K), the Keplerian prior (K$_{\rm p}$), 
kinematic constraint (K$_{\rm pk}$) and the Galactic prior  (K$_{\rm pkg}$) from the left to right panels, respectively.
The black, red, green and blue color-code correspond the models with the geometry of P$_{+-}$,  P$_{-+}$, P$_{++}$ and P$_{--}$,
respectively. 
The linear orbital motion models (L) are omitted for clarity, because they are almost same as the Keplerian orbit models (K).
}
\end{center}
\end{figure}

We first show the results with full Keplerian orbit (with scripts  ``K'') without any priors in Table \ref{tbl:param5} and Table \ref{tbl:physical5}.
The results are almost same as the ones with the linear approximation of the orbit. The large eccentricity of $e\sim$1 seeing 
the orbital plane face-on ($i\sim0,\, 180^{\circ}$ ) is as expected from the large ${\rm (KE / PE)_\perp}$ in the linear orbit. 
The physical parameter of the keplerian orbits, semi-major axis $a_{\rm kep}$, period $P$,  $e$, and $i$ 
are not well constrained so that they have very large asymmetric error bars in MCMC in Table \ref{tbl:physical5}.
Here, when the best fit is larger or smaller than the 68\% confidence interval of MCMC chains, 
the upper or lower limit is designated as ``$+0.0$'' or ``$-0.0$'', respectively.
So the light curve shape itself does not constrain the parameters more than the linear orbit model,
except that it ruled out the models with smaller $s_0$ and larger $ds/dt$ corresponding to ${\rm P_{\pm\pm}L'}$.
The best fit parallax vectors are larger than that of the static model as shown in Figure \ref{fig:piENE}.
Note that,  the ratio of 3D kinetic to potential energy ${\rm (KE / PE)}$ can be calculated 
in these full Keplerian orbit models as shown in Table \ref{tbl:physical5}. which are also close to unity. 

The results with the Keplerian orbit with the Keplerian prior (with scripts  ``K$_{\rm p}$'') are shown in 
Table \ref{tbl:param6}, Table \ref{tbl:physical6} and Figure \ref{fig:piENE}.
With the Keplerian prior, the circular orbits with $e\sim 0$ are preferred contrary to the large eccentricity 
without  the Keplerian prior.
This is because Jacobian $j_{\rm kep}$ is proportional to $1/(e\, \sin{i})$ as noted by \cite{Skowron2011} 
and thus smaller values of eccentricity are preferred.
Here, technically, the lower limit of eccentricity is set to be $10^{-4}$ to avoid a numerical problem in MCMC 
as suggested by  \cite{Skowron2011}.

$\pi_{\rm E}$ is reduced by a factor of 1/2$\sim$2/3 because the circular orbit is preferred by the Keplerian prior.
So the lens masses increased, while the hosts are still the high-mass and low-mass brown dwarfs. 
As for the models P$_{\pm\mp}$K$_{\rm p}$, there are other minima with a lower parallax value of $\pi_{\rm E} \sim 0.2$
with similar final $\chi^2$ whose host is a low-mass M-dwarf.
This is because $P_{\rm kep}$ prefers larger values of $D_l$ by $D_l^6$. 
But $\chi^2$ values from the light curves alone are larger than brown dwarf models. So there seems to be some conflict between 
light curve and prior.

 \subsubsection{Stellar Kinematic constraint}
\label{sec:kinematic}

\begin{figure}
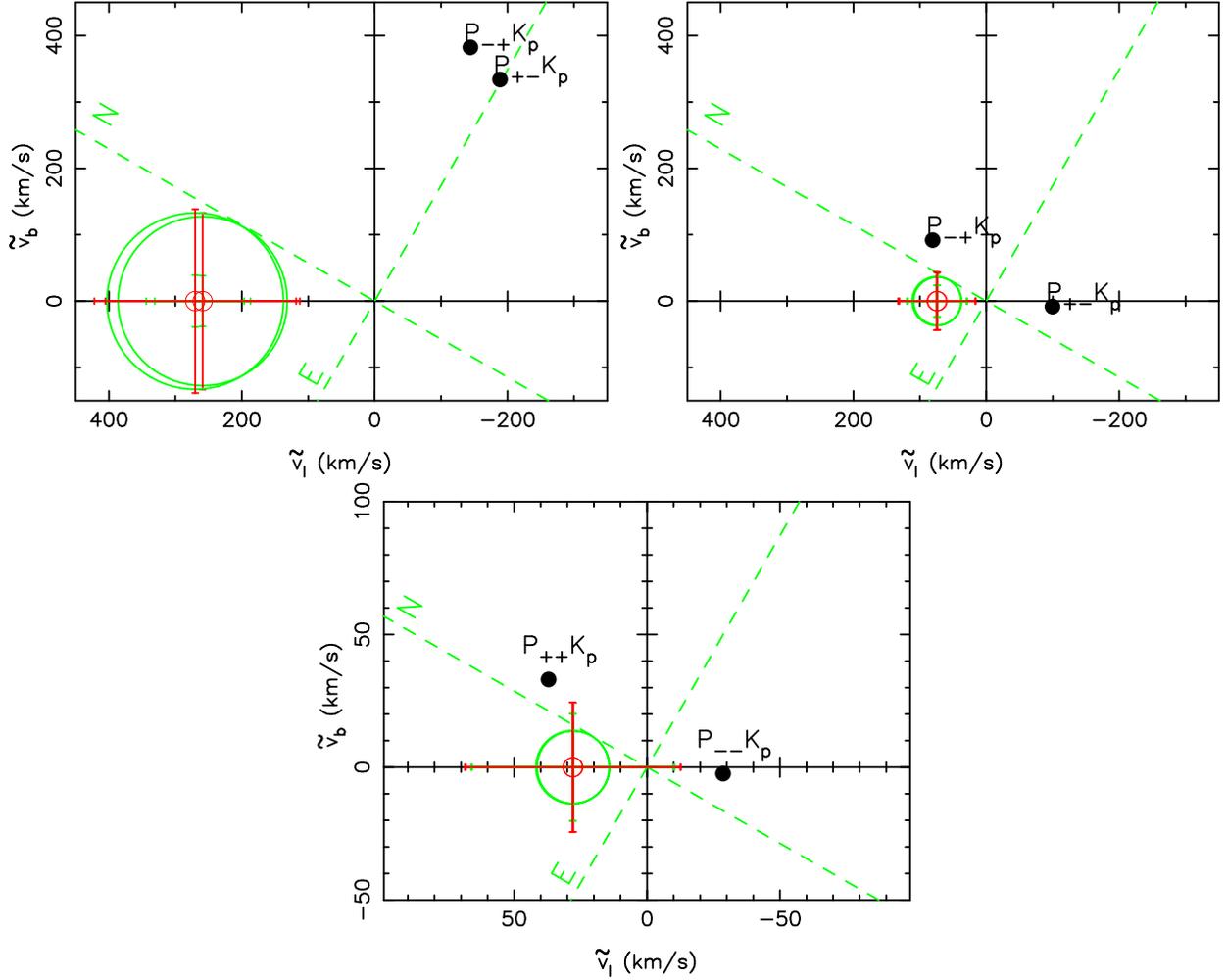

\begin{center}
\includegraphics[angle=-90,scale=0.39,keepaspectratio]{Figure8a.eps}
\includegraphics[angle=-90,scale=0.39,keepaspectratio]{Figure8b.eps}
\includegraphics[angle=-90,scale=0.39,keepaspectratio]{Figure8c.eps}
\caption{
  \label{fig:tilde_v}
Lens projected velocity \mbox{\boldmath $\tilde{v}_{\rm t}$} $=(\tilde{v}_{l}, \tilde{v}_{b})$ in the Galactic coordinate 
from the best fit with Keplerian orbit  with the Keplerian prior.
Top-left, top-right and bottom panels indicate M-dwarf and brown dwarf P$_{\pm\mp}$K$_{\rm p}$ models
and brown dwarf P$_{\pm\pm}$K$_{\rm p}$ models, respectively.
Red circles are expected mean from the Galactic kinematics.
Green error bars and circles represent uncertainty due to the velocity dispersion of disk stars and bulge stars, respectively.
They are added in the quadrature in the total error shown in red error bars. Green bashed line indicate the ecliptic North and East.
The  \mbox{\boldmath $\tilde{v}_{\rm t}$}  of M-dwarf P$_{\pm\mp}$K$_{\rm p}$ models are significantly different 
from the expected value from the Galactic kinematics.
}
\end{center}
\end{figure}

Here, we applied the prior for the Galactic kinematics by following \cite{Batista2011}.
In Table \ref{tbl:physical6}, the projected lens-source relative velocity \mbox{\boldmath $\tilde{v}_{\rm t}$} $=(\tilde{v}_{t,l}$, $\tilde{v}_{t,b})$ 
of these K$_{\rm p}$  models in the Galactic coordinate differs significantly. 
Those of the M-dwarf models are significantly different from the expected value from the Galactic kinematics
as shown in Fig. \ref{fig:tilde_v}.  Here we assume 
a source distance of $D_{\rm s} = 8$kpc 
 \citep{Reid1993,Honma2012},
the proper motion of the Galactic center is $\mu_{\rm GC} = 6.1$ mas\,yr$^{-1}$
 \citep{Backer1999,Reid2004,Honma2012} ,
the proper motion dispersion of stars in the bulge is $\sigma_{\mu,{\rm GB}} = 3$ mas\,yr$^{-1}$
 \citep{Kuijken2002},
the velocity dispersions of the Galactic disk stars in the Galactic coordinates are
$\sigma_{{\rm Disk},l}=34$ km\,s$^{-1}$ and $\sigma_{{\rm Disk},b}=18$ km\,s$^{-1}$
 \citep{Binney1998, Minchev2013,Sharma2014}.
Then the expected average $(\tilde{v}_{t,{\rm exp},l},\tilde{v}_{t,{\rm exp},b}) $ and 
dispersion $(\tilde{\sigma}_{t,l}, \tilde{\sigma}_{t,b})$ of the lens projected velocity are calculated.
The probability of having observed {\boldmath $\tilde{v}_{t}$}  can be given by

\begin{equation}
P_{\rm kin} = \exp{\left[  -\frac{(\tilde{v}_{t,l} - \tilde{v}_{t,{\rm exp},l})^2  }{2 \tilde{\sigma}_{t,l}^2 }    \right] }
                      \exp{\left[  -\frac{(\tilde{v}_{t,b} - \tilde{v}_{t,{\rm exp},b})^2  }{2\tilde{\sigma}_{t,b}^2 }    \right] }.
\label{eq:P_kin}
\end{equation}
The $\Delta\chi^2$ penalty of $\Delta \chi^2_{\rm kin} = -2 \ln(P_{\rm kin})$ is about +16 and +15 for 
M-dwarf P$_{+-}$K$_{\rm p}$ and P$_{-+}$K$_{\rm p}$  models respectively.
On the other hand the penalty is +9, +4, +2 and +2 for brown dwarf P$_{+-}$K$_{\rm p}$,
P$_{-+}$K$_{\rm p}$ , P$_{++}$K$_{\rm p}$  and P$_{--}$K$_{\rm p}$  models,
respectively. So the M-dwarf models are less preferred. 

Thus, we conducted MCMC runs by adding the penalty $\Delta \chi^2_{\rm kin}$.
The results (with scripts  ``K$_{\rm pk}$'')  are shown in Table \ref{tbl:param16}, Table \ref{tbl:physical16} and Figure \ref{fig:piENE}. 
The model light curve of P$_{++}$K$_{\rm pk}$ is shown in Figure \ref{fig:lcurve}.
As expected, $\pi_{\rm E}$ values for the M-dwarf models increase to $\sim 0.3$ to reduce the  $\Delta\chi^2_{\rm kin}$
and the total $\chi^2$ value became similar or larger than the brown dwarf models.  
In total, low-mass brown dwarf P$_{\pm\pm}$K$_{\rm pk}$ models are slightly preferred over other models.

We adopt these K$_{\rm pk}$ models as our main results of this paper
because they use the most realistic priors and constraints. See details in \S \ref{sec:model_selection}.

 \subsubsection{Galactic mass density prior}
\label{sec:galactic_prior}

Finally, we applied the prior for the Galactic mass density model \citep{Batista2011,Skowron2011}.

\begin{equation}
P_{\rm gal} = \nu(x,y,z)f(\mu)[g(M)M]\frac{D_l^4\mu^4}{\pi_{\rm E}},
\label{eq:P_gal}
\end{equation}
which is the microlensing event rate multiplied by the Jacobian of the transformation 
from microlensing parameters to physical coordinates,
$j_{\rm gal}=||\partial (D_l, M, \mbox{\boldmath $\mu$})/ \partial (t_{\rm E}, \theta_{\rm E},  \mbox{\boldmath $\pi_{\rm E}$})||$.
Here $ \nu(x,y,z)$ is the local density of lenses, $g(M)$ is the mass function.  
$f(\mu)$ is the two-dimensional probability function for a given source-lens relative proper motion, 
$\mu=v_{\rm t}/D_l$, which is set to unity because it is already implemented in $P_{\rm kin}$ above.
We adopt the Galactic model by \cite{han95} for $ \nu(x,y,z)$ and 
adopt $g(M)\propto M^{-1}$ by following  \cite{Batista2011}.

The results of MCMC runs by adding a penalty of $\Delta \chi^2_{\rm gal} = -2 \ln(P_{\rm gal})$  
 are shown (with scripts  ``K$_{\rm pkg}$'')  in Table \ref{tbl:param26}, Table \ref{tbl:physical26} and Figure \ref{fig:piENE}.
 The light curves,  caustics, critical curves and source trajectory of the models comprising both parallax and planetary 
orbital motion with various different priors are almost same as that of the parallax-only models as shown in Figure \ref{fig:lcurve}
 and Figure \ref{fig:caustic}.
Overall, the lens masses slightly increased relative to that of K$_{\rm pk}$ models because this P$_{\rm gal}$ 
prefers larger $D_l$, i.e., smaller $\pi_{\rm E}$.

In addition to above six models, there are two more minima for P$_{\pm\pm}$K$_{\rm pkg}$ 
with a lower parallax value of $\pi_{\rm E} \sim 0.8$ with similar final $\chi^2$ values.
This is also because the prior P$_{\rm gal}$ prefers larger $D_l$ and smaller $\pi_{\rm E}$ values. 
These solutions happen to have similar $\pi_{\rm E}$ values with that of high-mass brown dwarf P$_{\pm\mp}$K$_{\rm pkg}$ models,
hence the similar physical parameters.

As for the P$_{\pm\mp}$K$_{\rm pkg}$ models, there are two other minima
which each have a  much smaller parallax value of $\pi_{\rm E} \sim 0.035$ and a smaller final $\chi^2=$15107. 
This is because their large values of  $D_l=7$\,kpc  are preferred by $P_{\rm gal}$.
These solutions have a very heavy host mass of $M_{\rm h}\sim1.7M_\sun$ which would be quite rare.
 In addition, these solutions have a value of $\chi^2_{\rm lc}\sim 15224$ which is larger than any other model with parallax,
which conflict to the preference by P$_{\rm gal}$.
Furthermore, these models have very bright $H$-band source magnitudes  $H_{\rm s} =  15.816 \pm 0.017$
and $H_{\rm s} =  15.824 \pm 0.018$ for P$_{++}$K$_{\rm pkg}$ and P$_{--}$K$_{\rm pkg}$, respectively.
These are too bright compared to  the Keck AO measurement of the target, $H=15.90 \pm 0.02$, by  3$\sigma$
(see Section \ref{sec:KeckAO}). 
If we assume that the host is a main sequence star, then total brightness of the source plus lens 
is expected to be brighter and ruled out by Keck measurements by more than 4$\sigma$.
For these reasons we do not consider these solutions to be real, and are not listed amongst the other solutions in the Tables.


 \subsection{Model Selection}
\label{sec:model_selection}

We have presented a large number of models with various high order effects and
priors. Here we summarize which of these models are preferred over the others.
As discussed in \S \ref{sec:parallax}, the parallax signal looks qualitatively real because
the signal come from the theoretically expected part of the light curve and it is consistent 
in both MOA and OGLE-$I$ datasets. 
The $\chi^2$ improvement by the parallax-only models over the standard model are 
$\Delta\chi^2=-33.3\sim-37.1$ with 2 additional parameters, which is equivalent  to a confidence level 
of 5.4$\sim$5.8$\sigma$ and is formally significant. 

Furthermore, we compared the models by using the common statistical criteria, 
Akaike's Information Criterion (AIC),  $AIC= \chi^2 + 2n_{\rm param}$,  and
the Bayesian information criterion (BIC), $BIC = \chi^2 + n_{\rm param} \ln(N_{\rm data}$) \citep{Burnham2002},
which includes a penalty discouraging an overfitting. The smaller the $AIC$ and/or $BIC$ values are the better model is.
Here we adopt the number of data points $N_{\rm data}=2913$ during the event at $6450<HJD'< 6620$
for $BIC$ because the baseline data outside of this range do not constrain the parallax signal.

The differences in these criteria between the parallax-only models and standard model are $\Delta AIC= -29 \sim -33$ and
$\Delta BIC= -17\sim -21$.  Thus the parallax-only models are better than the standard model.
The parallax models with linear orbit ($L$) are also better than standard model by $\Delta AIC= -28 \sim -41$ and
$\Delta BIC= -5 \sim -17$. 
The parallax models with full keplerian orbit with two more parameters are better than the standard in $AIC$ but not in $BIC$.

The $\Delta\chi^2$ of the parallax models with linear orbit ($L$) relative to the parallax-only models are 
only marginal of $-3\sim-12$ with two additional parameters.  
The differences in these criteria are $\Delta AIC= +0.7 \sim -8$ and
$\Delta BIC= +13 \sim +4$, and it is worse for the full keplerian orbit with two more parameters.
Thus, the inclusion of the orbital motion is not justified by these criteria.

However, the reason that we must introduce the orbital motion is not to improve the goodness of the fit, but
to avoid the bias of the value and the underestimate of the uncertainty in parallax parameters due to the known degeneracy 
between the parallax and the lens orbital motion as shown in section \S \ref{sec:LinearOrbitalMotion}.

Furthermore, even though the full Keplerian orbit does not improve the goodness of 
the fit with two additional parametes, its incorporation has the following benefits.
As Skowron et al (2011) noted, in addition to being more accurate, the advantage of 
the full Keplerian orbit is to avoid all unbound orbital solutions (with eccentricity $>1$) 
and to enable the introduction of priors on the values of orbital parameters directly 
into MCMC calculations. 
If the uncertainty of the parameters are relatively large like in this event, the density 
distribution of the MCMC chain depends on the prior probability of the fitting parameters
in MCMC.  However, the linear orbital motion parameterization inadvertently assumed 
the uniform prior on all microlensing fitting parameters, which is not physically justified.  
On the other hand, the full keplerian orbit parameterization enable us to properly weigh 
the MCMC chains with physically justified priors.

We conducted the modeling with three different sets of relatively realistic priors including the Keplerian prior, i.e.,
K$_{\rm p}$, K$_{\rm pk}$, and K$_{\rm pkg}$. 
We think that the model K$_{\rm pk}$ is more realistic than K$_{\rm p}$ as the galactic kinematics constraint is applied.
The models with the Galactic mass density prior K$_{\rm pkg}$ may also be useful, but
we do not know if the assumption that the distribution of the planetary systems is uniform throughout the Galaxy is valid.
Thus we adopt the K$_{\rm pk}$ models as our main results of this paper. 

It is important to note that the results of all these models K$_{\rm p}$, K$_{\rm pk}$, and K$_{\rm pkg}$,
are basically same within their errors, thus  our main conclusion does not depend on 
the choice of these priors.

Among the models in K$_{\rm pk}$, the low-mass brown dwarf models, 
i.e., P$_{++}$K$_{\rm pk}$ and P$_{--}$K$_{\rm pk}$, are slightly 
preferred by both $\chi^2_{lc}$, $\chi^2$ and $H_{\rm s}$ (see \S\ref{sec:KeckAO}). 
But we accept all the modes equally as possible solutions.

\section{Lens properties}
\label{sec:property}
The lens physical parameters can be derived for this event because we could measure 
both the parallax and finite source effects in the light curve.

\begin{figure}
\begin{center}
\includegraphics[angle=0,scale=0.7,keepaspectratio]{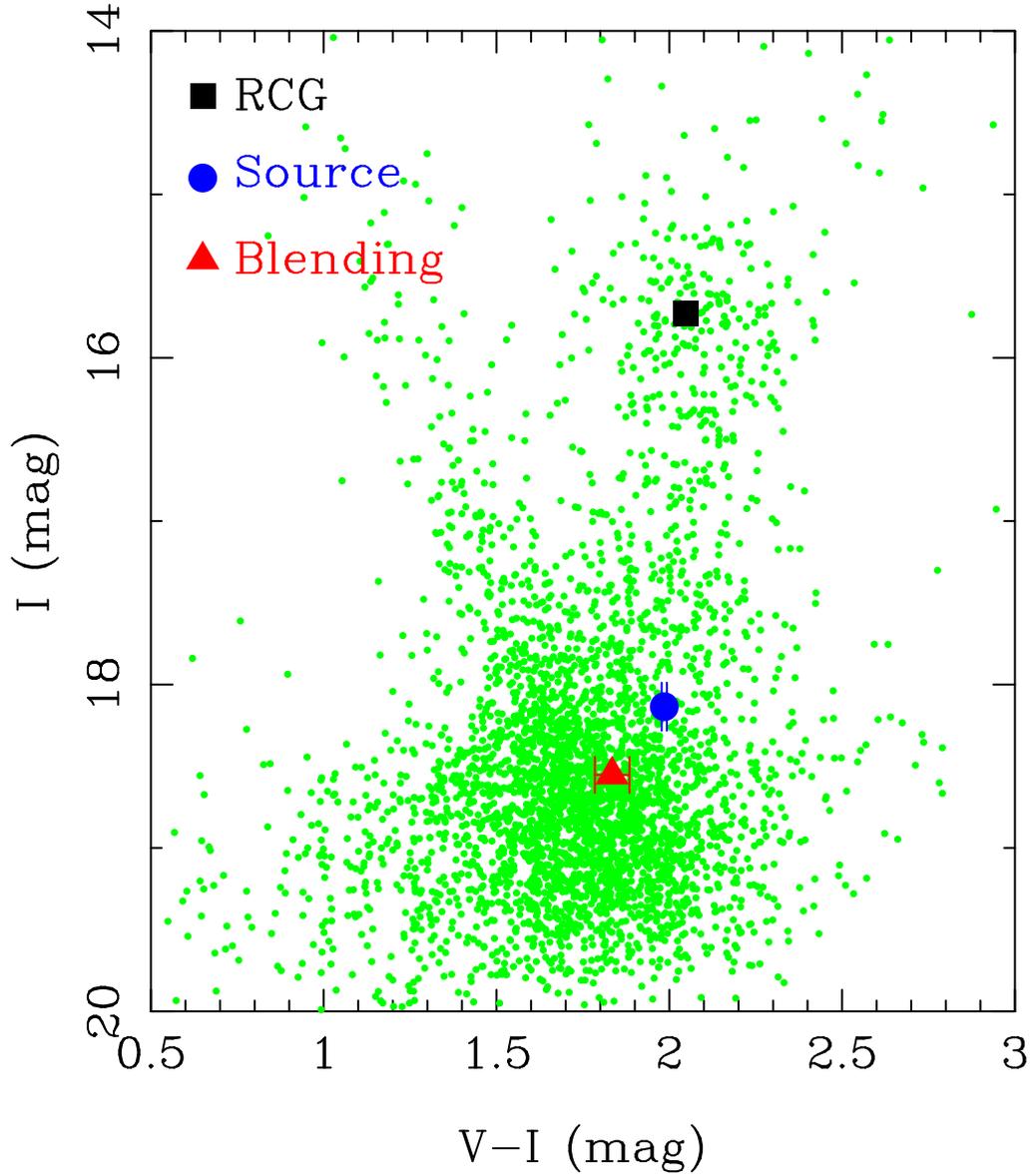}  %
 \caption{
  \label{fig:cmd}
OGLE-IV calibrated $V$ and $I$ Color magnitude diagram (CMD) within 2'$\times$2'
around the event (green dots). A filed square indicates the center of
Red Clump Giants.  The source position of P$_{++}$K$_{\rm pk}$ model 
shown by the filled circle, which is almost identical for all models,  
indicates that the source star is a K2 sub-giant.  
}
\end{center}
\end{figure}

The OGLE-IV calibrated color magnitude diagram (CMD) in a $2'\times2'$ region around the event is 
shown in Figure \ref{fig:cmd}. Figure \ref{fig:cmd} also shows the center of the Red Clump giants (RCGs)
$(V-I,I)_{\rm RC,obs} =( 2.047, 15.73) \pm (0.002, 0.04)$ and
the model independent OGLE $V-I$ source color found by linear regression and the best fit source $I$ magnitude 
of the model P$_{++}$K$_{\rm pk}$,  $(V-I,I)_{\rm s}=( 1.985, 18.13 ) \pm (0.008, 0.02)$.
 $I_{\rm s}$ for other models are almost same, as shown in Table \ref{tbl:param}-\ref{tbl:param26}.

Assuming the source suffers the same dust extinction and reddening as the RCGs and using the
expected extinction-free RCG centroid
$(V-I,I)_{\rm RC,0}   =( 1.06, 14.39) \pm (0.06,0.04)$ at this position
\citep{Bensby2013,Nataf2013}, we estimated the extinction-free color and magnitude of the source as
$(V-I,I)_{\rm s,0}    =( 1.00 , 16.80 ) \pm (0.06, 0.06)$.  This color measurement is consistent with the independent 
measurement of $(V-I)_{s,0}=1.02\pm0.06$ by the CTIO telescope  (see \S \ref{sec:data}).
We use the average of OGLE and CTIO colors, $(V-I)_{\rm s,0} = 1.01  \pm 0.06$ in the following analyses. 
Here the errors in $(V-I)_{\rm s,0}$ are dominated by the error in $(V-I)_{\rm RC,0}$.
These values are consistent with the source being a K2 subgiant \citep{bessell88}.


Following \cite{Fukui2015}, we estimated the source angular radius, $\theta_*$, by using the relation between
the limb-darkened stellar angular diameter, $\theta_{\rm LD}$, $(V-I)$ and $I$ given by Equation (4) of \cite{Fukui2015}. 
This relation is derived from a subset of the interferometrically measured stellar radii in \cite{Boyajian2014}, in which 
the dispersion of the relation is $\sim$2\% by using only stars with $3900 K < T_{\rm eff} < 7000$ K to improve the fit for FGK stars. 
This yield the source radius of $\theta_*=\theta_{\rm LD}/2= 1.84\pm 0.12\,\mu$as.


The spectrum of the source was taken by the UVES spectrograph on the Very Large Telescope (VLT) 
at a time when  the source was still magnified as a Target of Opportunity (ToO) observation. 
Reductions were carried out with the UVES pipeline \citep{Ballester2000}.
The observation and data analysis have been done by the same manor as \cite{Bensby2011,Bensby2013}.
This gives 
the source effective temperature, $T_{\rm eff} = 4854 \pm 66$ K,  the gravity, $\log g = 3.30 \pm 0.14$, 
and the metallicity, [Fe/H]$ = -0.17 \pm 0.09$ (Bensby et al. in preparation).
By using these values and the  the relation by \cite{Casagrande2010}, we  derive the extinction-free source colors, 
$(V-I)_{\rm s,0,spec}   = 1.036 \pm 0.047$ and $(V-H)_{\rm s,0,spec}   = 2.244 \pm 0.078$. 
Thus we get $(I-H)_{\rm s,0,spec}= 1.208 \pm 0.091$.
The extinction-free $H$-band source magnitude is given as $H_{\rm s,0} =  I_{\rm s,0}-(I-H)_{\rm s,0,spec}=15.59 \pm 0.11$.

The expected extinction-free $(I-H)_{\rm s,0}$ from the measured $(V-I)_{\rm s,0}$ are
$(I-H)_{\rm s,0} = 1.188 \pm 0.082$ and
$(I-H)_{\rm s,0} = 1.119 \pm 0.074$ by using the stellar color-color relation of  
\cite{bessell88} and \cite{Kenyon1995}, respectively.
These are roughly consistent to $(I-H)_{\rm s,0,spec}$.
By a linear regression of OGLE-$I$-band light curve and CTIO $H$-band light curve (see \S \ref{sec:data}), 
which are calibrated to the 2MASS scale, we got the source $(I-H)$ color as, 
\begin{equation}
(I- H)_{\rm s} = 2.256 \pm 0.016.
\label{eq:IHs}
\end{equation}
Correcting the extinction by using the measured extinction and reddening $(E(V-I), A_I)$
by RCGs and the extinction law of \cite{Chen2013}, we got
$(I-H)_{\rm s,0,OGLE,CITO} = 1.259 \pm 0.071$,
which is also consistent with $(V-H)_{\rm s,0,spec}$.

Then we got  $\theta_*= 1.90\pm 0.11\,\mu$as by using  the relation between $\theta_{\rm LD}$, $H_{s,0}$,  $(V-H)_{\rm s,0}$ and [Fe/H],
given by Equation (9) of \cite{Fukui2015}, which is also driven in the same way as  Equation (4) of \cite{Fukui2015}
but with the metallicity term. Note $H$ in the relation is in Johnson magnitude system. Thus the observed $H$-band 
source magnitude which is in the 2MASS system, is converted to the Johnson system by following \cite{Fukui2015}. 
This is consistent with above value.
The average of above values are,
\begin{equation}
\theta_*= 1.87\pm 0.12\,\mu {\rm as},
\label{eq:thetas}
\end{equation}
where we adopt the larger error from the estimate with $(V-I,I)$, conservatively.
This value is about the median of those from other models and differences from them are less than 2\%, 
thus we adopt this value for all models in the following analysis.

We also tested the traditional method as follows.
Following \cite{yoo04}, the dereddened source color and brightness
($V$-$K$, $K$)$_{\rm s,0}=(2.2, 15.6)$ are estimated using the observed
($V$-$I$, $I$)$_{\rm s,0}$ and the color-color relation of
\cite{Kenyon1995}. 
By using this ($V$-$K$, $K$)$_{\rm s,0}$ relation and the empirical color/brightness-radius relation of
\cite{kervella2004},
we estimated the source angular radius, $\theta_*= 1.85 \pm 0.16  \,\mu$as, where
the error includes uncertainties in the color conversion and the color/brightness-radius
relations. This is consistent with the above value.

The physical parameters of all models are listed in Table \ref{tbl:physical}-\ref{tbl:physical26}.
The physical properties of three models with realistic priors and constraints, i.e., K$_{\rm p}$, K$_{\rm pk}$, and K$_{\rm pkg}$, 
are basically same within the error bars.
In the following analysis, we focus on the model K$_{\rm pk}$.

Here we summarize the physical properties of the lens system by showing the average values of various models in Table \ref{tbl:physical16} for clarity.
The averages are taken without any weighting by their error bars. 
The uncertainties are given by the maximum and minimum values of 1-$\sigma$ upper and lower limits of all (or a group of) models.
The angular Einstein radii, and geocentric lens-source relative proper motion $\mu_{\rm geo}$, 
which are independent of 
the parallax values, 
are estimated, respectively, as follows,

\begin{eqnarray}
\theta_{\rm E}&=& {\theta_*\over\rho} =    0.48\pm 0.06   \, {\rm mas},  \label{eq:thetaE}\\ 
\mu_{\rm geo}&=&  {\theta_E \over t_{\rm E} } =8.4\pm1.2   \, \rm mas\,yr^{-1}. 
\label{eq:mu}
\end{eqnarray}
This  $\mu_{\rm geo}$ is consistent with the typical value for disk lenses of $\mu\sim$5-10 $ {\rm mas\,yr^{-1}}$
\citep{han95}.

The total mass and distance of the lens system can be given by $M=\theta_{\rm E}/(\kappa \pi_{\rm E})$ and
$D_{\rm l}=\rm AU/(\pi_{\rm E}\theta_{\rm E}+\pi_{\rm s})$,
where $\kappa =4G/(c^2$AU)=8.144 mas $M_\sun^{-1}$,  $\pi_{\rm s}=$AU/$D_{\rm s}$ 
and $D_{\rm s}\sim8$ kpc is the distance to the source 
\citep{Dong2009-MOA400}.
Thus these quantities depend on the parallax parameter and we have 
three groups of solutions in models, i.e, 
small $\pi_{\rm E}\sim0.3$ (P$_{\pm\mp}$K$_{\rm pk}$), medium $\pi_{\rm E}\sim0.8$ (P$_{\pm\mp}$K$_{\rm pk}$ and 
P$_{\pm\pm}$K$_{\rm pk}$) and large $\pi_{\rm E}\sim2$ (P$_{\pm\pm}$K$_{\rm pk}$).
The distance to the system, $D_{\rm l}$, the mass of the host, $M_{\rm h}$, and planet, $M_{\rm p}$, and 
their projected separation, $a_\perp$, of these solutions are,

\begin{equation}
D_{\rm l}=3.6_{-0.8}^{+0.6}\, {\rm kpc},\,\,  1.8_{-0.2}^{+0.4}\, {\rm kpc},\, {\rm or}\,\, 0.85_{-0.08}^{+0.13}\, {\rm kpc},
\label{eq:Dl}
\end{equation}

\begin{equation}
M_{\rm h}=\frac{M}{1+q} =0.19_{-0.06}^{+0.05}\, M_\sun,\,\,  0.068_{-0.011}^{+0.019}\, M_\sun,\, {\rm or}\,\, 0.025_{-0.004}^{+0.005}\, M_\sun,
\label{eq:mass_h}
\end{equation}

\begin{equation}
M_{\rm p}=\frac{qM}{1+q} =21_{-7}^{+6}  M_\Earth,\,\, 7.9_{-1.2}^{+1.8}  M_\Earth,\, {\rm or}\,\, 3.2_{-0.3}^{+0.5} M_\Earth, 
\label{eq:mass_p}
\end{equation}

\begin{equation}
a_\perp=s\theta_{\rm E}D_{\rm l}=4.2_{-0.9}^{+0.7} \, {\rm AU},\,\,   2.1_{-0.2}^{+0.4}\, {\rm AU}, {\rm or} \,\, 0.94_{-0.09}^{+0.12}  \, {\rm AU},
\label{eq:aproj}
\end{equation}
respectively. 
Here $a_\perp$ is the 2-dimensional (2D) projection of a 3D elliptical orbit having a semi-major axis $a$.
The expected 3D semi-major axis can be estimated by $a_{\rm exp}=\sqrt{3/2}a_\perp$
 \citep{Gould2014}.
The best fit 3D semi-major axis by the Keplerian orbit, $a_{\rm kep}$, are in-between of these values in 
the case of this event.

 The semi-major axis $a_{\rm kep}$ normalized by the snow-line, $a_{\rm snow}=2.7(M_h/M_\sun)$, are

\begin{equation}
\frac{a_{\rm kep}}{a_{\rm snow}} = 8.9_{-1.4}^{+10.5},\,\,  12_{-1}^{+7},\,\, {\rm or}\,\, 14_{-1}^{+11},
\label{eq:asnow}
\end{equation}
The effective temperature of the planet at the time of its formation based on the host mass
 and host-planet separation are also given in the tables.
 
The small parallax models suggest that the planet  has a mass similar to Neptune  (17M$_\Earth$) 
orbiting  a very low mass M-dwarf in the Galactic disk.
The planet is very cold as the estimated separation is $8.9_{-1.4}^{+10.5}$ times larger than the snow-line.
This is comparable to Neptune's semi-major axis, i.e., 11 times larger than the Sun's snow-line.
This interpretation of the planetary signal for MOA-2013-BLG-605Lb, therefore, suggests the planet is a Neptune analog.

The medium parallax models correspond to a miniature Neptune (or large-mass ``super-Earth'') orbiting a high-mass brown dwarf host.
The planet is even colder as the planetary orbit radius is $12_{-2}^{+7}$ times larger than the 
snow line, which is also similar to the Neptune.

The large parallax models correspond to 
three times the Earth-mass planet orbiting a low-mass brown dwarf host.
The planet is colder because the planetary orbit radius is $14_{-1}^{+11}$ times larger than the 
snow line.

These solutions of K$_{\rm pk}$ are compared to the planets found by other methods in Figure \ref{fig:a_Mp}.
As one can see in the right-hand panel of Figure \ref{fig:a_Mp}, in either group of models, this planet is 
the coldest low mass planet ever found and it is very similar to Neptune.


\section{Keck AO Observations and Lens Mass constraint}
\label{sec:KeckAO}

\begin{figure}
\begin{center}
\includegraphics[angle=0,scale=0.6,keepaspectratio]{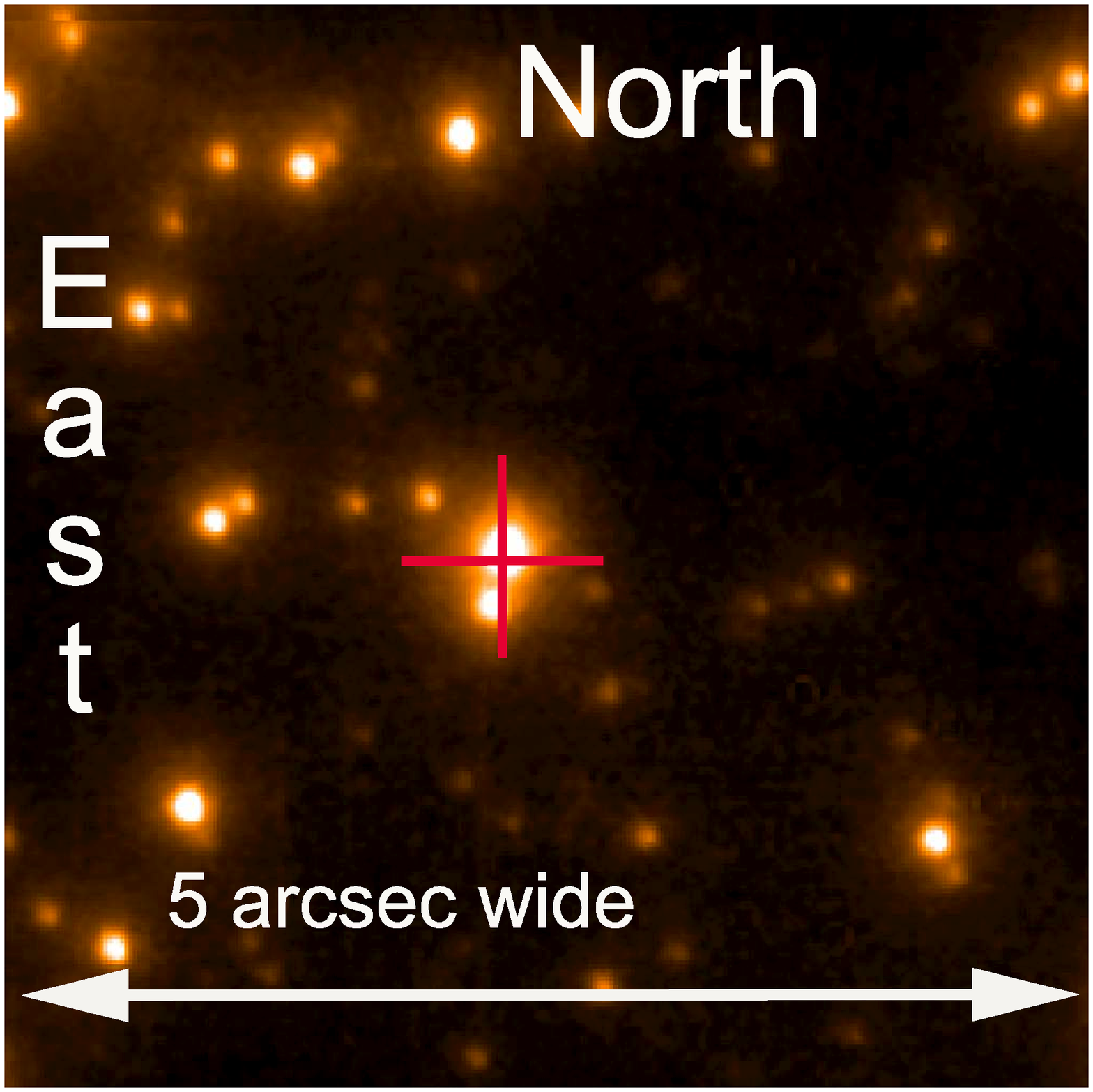}
\caption{
  \label{fig:KeckAO}
$H$-band Keck AO image within 5"$\times$5" around the event.
The red cross indicates the actual source position measured during the magnification 
on the OGLE difference image.
The brighter star at the cross is the source with $H=15.90 \pm 0.02$ mag. 
The fainter star on South is  the blend with $H=17.01 \pm 0.02$ mag. 
This shows that the blend measured in the fitting process is not the lens. 
The measured $H$-band flux of the source shows 
that the lens is very faint and not detected.
}
\end{center}
\end{figure}

We observed with the NIRC2 instrument mounted on KECK-II the microlensing target MOA-2013-BLG-0605 on July 26, 2015. 
We used the Wide camera giving a pixel scale of 0.04 arcsec and a field of view of 40 arcsec. 
We adopted a 5 position dithering pattern, and did 30 exposures of 10 seconds each. We performed dark subtraction 
and flatfielding in the standard manner for an IR detector. We then stacked the frames using Swarp \citep{Bertin2002}, without subtracting the background. The final 
image is shown in Figure \ref{fig:KeckAO}.

For absolute calibration, we used images from the VVV survey done with the VISTA 4m telescope at Paranal \citep{Minniti2010}.
We extracted a 3 arcmin JHK band images centred on the target. We computed a PSF model using PSFEX software \citep{Bertin1996}, and measured
fluxes on the frames using SExtractor with this PSF model. We cross identified the stars from the field with 2MASS catalogues. 
 We selected 300 stars that are bright while not saturated on VVV, and derive the photometric zero points with an accuracy of 0.004 mag.
We then use the VVV catalogue to perform the astrometric calibration of the KECK frame. 

We then measure the fluxes using SExtractor as described in  \citep{Batista2014}. We cross identified 39 stars in both the VVV 
and KECK image. We exclude the stars saturated on the KECK, and derive the zero point of KECK photometry. 
In the non-AO PSF, there are two stars, the source with $H=15.90 \pm 0.02$ and a blend at $\sim 0.3$\,arcsec 
to the south with $H=17.01 \pm 0.03$.
Here, these two stars are blended in the OGLE reference image and the cataloged centroid is in-between of them. 
The actual source position during the magnification on the OGLE difference image was precisely measured 
as shown by the red cross in Figure \ref{fig:KeckAO}.
This clearly resolved the source and showed that the blend measured in the fitting process is not the lens.

By using the apparent source color $(I- H)_{\rm s}$ given by Eq. (\ref{eq:IHs}) 
and the best fit $I_{\rm s}$, $H$-band source magnitude, $H_{\rm s}$ are calculated 
as shown in Table \ref{tbl:physical}-\ref{tbl:physical26}.
There is a trend that the smaller the parallax is, the brighter the source is.
The $H_{\rm s}$ of low-mass brown dwarf models are almost same as the Keck measurement of $H=15.90 \pm 0.02$ 
or only slightly brighter within 1$\sigma$. The $H_{\rm s}$ of high-mass brown dwarf and most of M-dwarf models are within 2$\sigma$.
The M-dwarf models, P$_{-+}$K$_{\rm p}$, P$_{-+}$K$_{\rm pk}$  and P$_{-+}$K$_{\rm pkg}$ have
$H_{\rm s}=15.856\pm0.022$,
                 $15.856\pm0.024$,
         and  $15.845\pm0.026$, which are  1.6, 1.7 and 2.1 $\sigma$ brighter than the Keck measurement.
Furthermore, these are 2.0, 2.0 and 2.4 $\sigma$ brighter when those include the lens (host) brightness (\citealt{Kroupa1997}) of 
$H_{\rm h} =  20.96 \pm  0.24$ ($M_{\rm h}=0.28M_\sun$, $D_l=4.3$kpc),  
                     $21.22 \pm  0.24$ ($M_{\rm h}=0.20M_\sun$, $D_l=3.5$kpc) 
            and  $21.17 \pm  0.24$ ($M_{\rm h}=0.21M_\sun$, $D_l=3.7$kpc), respectively.
These comparisons of the Keck result and source magnitudes from the light curve indicate 
that the lens is very faint and not detected.
This is consistent with the physical solutions from the parallax measurements that the host 
is a low-mass M-dwarf or a brown dwarf mentioned in Section \ref{sec:property}. 
The low-mass brown dwarf models are slightly preferred.
The $H_{\rm s}$ of the standard model is consistent to the Keck results, but we concluded 
that the parallax models are better as discussed in \S \ref{sec:model_selection}.

As mentioned in Section \ref{sec:galactic_prior}, there are two minima with a much smaller parallax value of $\pi_{\rm E} \sim 0.035$ 
for the P$_{\pm\mp}$K$_{\rm pkg}$ models.
In addition to the rarity  of their heavy host mass of $\sim1.7M_\sun$ which might be a stellar remnant,
their source magnitudes 
$H_{\rm s}=15.824\pm0.018$ and
$H_{\rm s}=15.816\pm0.017$ are 2.9$\sigma$ and 3.2$\sigma$ brighter than the Keck measurement.
So these models are not likely real.
If their host is a main sequence star, then total brightness of the source plus lens 
are ruled out by Keck measurement by more than 4$\sigma$.


\section{Future Mass Measurement}
\label{sec:future}
Let us consider the prospects for resolving the degeneracy and characterizing the host and the planet.
In the first epoch of Keck AO observations, we could not detect any excess light, which 
confirmed the lens is faint.
If the second epoch is taken by {\it HST} or AO observations, then we may directly detect the host (or possibly its companion).
We can then measure the lens mass and distance or place a stronger upper limit on the lens mass. 

In Table \ref{tbl:physical}-\ref{tbl:physical26}, the geocentric proper motions are reported as
 $|\Vec{\mu}_{\rm  geo}| = \theta_{\rm E} / t_{\rm E} = 8\sim 9$ \,mas yr$^{-1}$. 
 The heliocentric proper motion is given by \citep{Janczak2010},

\begin{equation}
\Vec{\mu}_{\rm hel} = \Vec{\mu}_{\rm geo} + \Vec{v}_{\oplus,\perp} \frac{\pi_{\rm rel}}{{\rm AU}},
\label{eq:mu_hel}
\end{equation}
where   $\Vec{v}_{\oplus,\perp}=(v_{\oplus,\perp,N}, v_{\oplus,\perp,E})=(-2.96, -8.24)$ km\,s$^{-1}$ is the velocity of the Earth projected on the plane of the sky at the peak of the event. 
The estimated $\Vec{\mu}_{\rm hel}=(\mu_{\rm hel, N},  \mu_{\rm hel, E})$ of each model
is shown in Table  \ref{tbl:physical}-\ref{tbl:physical26},
 and they are about $8\sim 9$ mas\,yr$^{-1}$.
Hence it is clear that the lens will be separately resolved by {\it HST} or AO observations in 5-10 years' time given a diffraction limit of 50 mas.
Or, if we do not see any luminous object, then the lens is a sub-stellar object.

Not only the value but also the direction of expected relative proper motion would help us to know 
if it was the lens or  just an ambient star when we detect such star at 80mas from the source 10 years later.





\section{Discussion and Conclusion}
\label{sec:discussionAndSummary}

There are three physical planetary solutions for the MOA-2013-BLG-605L system.
One comprises a Neptune-mass planet at a wide separation from a very low mass 
M-dwarf host star, having a very similar temperature as Neptune when the planet was formed.
The second solution comprises a mini-Neptune around a high-mass brown dwarf which is 
even colder than Neptune when it was formed.
The third one is a super-Earth around a low-mass brown dwarf.

These degenerate solutions may be resolved by future high resolution imaging of 
the lens by the {\it HST} or ground-based  telescopes using adaptive optics, after waiting a period of time for the positions of the lens and the source to diverge. We may detect an M-dwarf lens host star, but we do not expect to detect a brown dwarf host star by such direct imaging.

In either case, the host is one of the three least massive main sequence 
stars orbited by a planet for which the planet's mass was measured and for which the planet-host mass ratio is $q<0.01$. The other low host mass, low planet mass systems are
MOA-2007-BLG-192L ($M_{\rm h}= 0.084_{ -0.012}^{+0.015}M_\sun$, 
$M_{\rm p}=3.2_{-1.8}^{+5.2} M_\earth$
$r_\perp=0.66_{-0.22}^{+0.51}$ AU) \citep{bennett08,kubas2012},
MOA-2010-BLG-328L
($M_{\rm h}= 0.11\pm 0.01M_\sun,
M_{\rm p}= 9.2 \pm2.2M_\earth,
r_\perp = 0.92\pm0.16$\,AU)
\citep{furusawa2013}.

These planets found around very low mass ($\sim 0.1 M_\sun$) hosts have 
relatively small masses themselves, ranging from super-Earth mass to Neptune mass. In contrast, a roughly equal number of giant planets and planets with Neptune-mass or less 
have been found across the whole mass range of host stars.
This may imply that the formation of gas giants is more difficult around very low mass stars
compared to average K-M dwarf stars with masses of $\sim0.5 M_\sun$, which is the typical host star for microlensing planets. This is somewhat as predicted by the core accretion model of planetary formation,
but this work provides the  first observational evidence supporting this prediction.

This could be the first exoplanet around a brown dwarf with a mass measurement 
having a  planetary mass-ratio $q<0.03$.
 There are three brown dwarf binaries where one 
of the components is in the planetary mass regime \citep{choi2013,han2013}. However, their 
mass ratios are large $q\ge0.08$, suggesting that their formation may be considered more akin to binary formation than planetary formation.

\begin{figure}
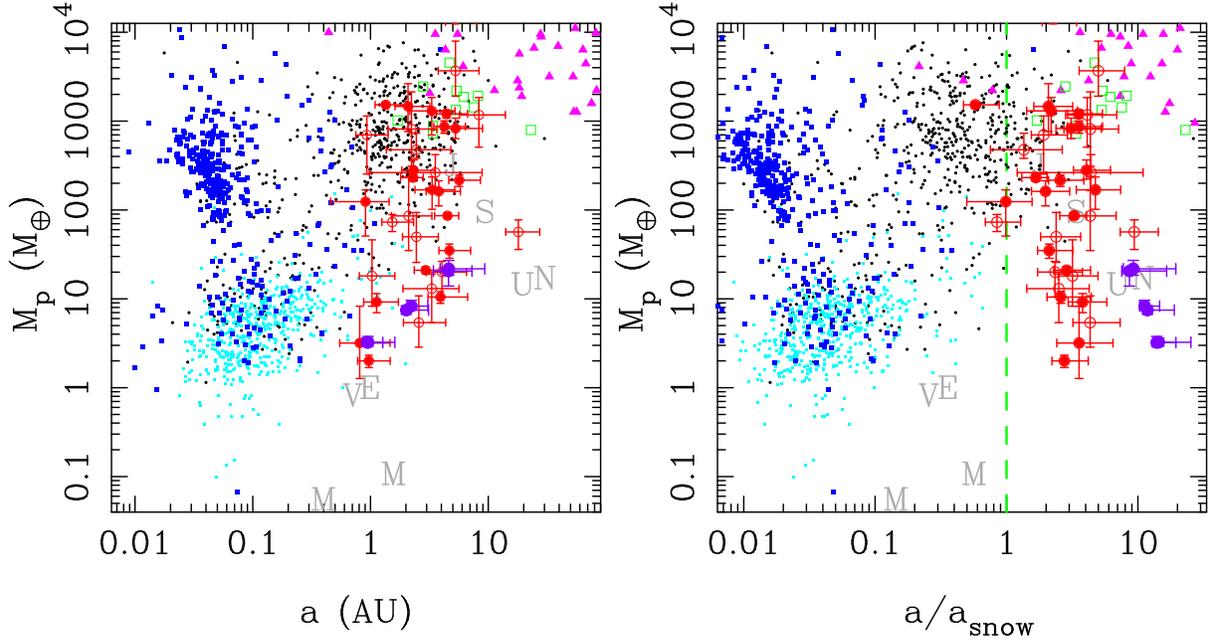

\begin{center}
\includegraphics[angle=-90,scale=0.5,keepaspectratio]{Figure1a.eps} 
\includegraphics[angle=-90,scale=0.5,keepaspectratio]{Figure1b.eps} 
\caption{
  \label{fig:a_Mp}
The distribution in planetary mass, $M_{\rm p}$, versus 
the semi-major axis, $a$ (left panel) and
 $a$ normalized by the snow-line (right panel) 
of discovered exoplanets by various methods.
Red circles indicate the microlensing planets. Microlensing planets for which mass measurements have been made are indicated with filled circles. 
Microlensing planets where the mass has been estimated by a Bayesian analysis  are indicated with open circles.
The six model solutions for event MOA-2013-BLG-605Lb comprising parallax and the Keplerian orbital motion 
with the Keplerian prior and the kinematic constraint (K$_{\rm pk}$)
are indicated by purple filled circles.
Black dots represent the radial velocity planets and blue filled squares 
are transit planets. Cyan dots are transit planets found by \emph{Kepler}. 
Magenta triangles denote planets found via direct imaging.
Green open squares denotes planets found via timing measurements. Solar system planets are indicated by their initial.
A green vertical dashed line indicates the snow line.
All models for MOA-2013-BLG-605Lb  are very similar to Neptune, when planet orbit radii are scaled to the snow line (right panel).
}
\end{center}
\end{figure}

The separation of the planet is very wide, 
 $8.9_{-1.4}^{+10.5}$,  $12_{-1}^{+7}$ or $14_{-1}^{+11}$  times larger than the snow line
 of $\sim$0.5($M /0.2M_\sun$) AU,  $\sim$0.2($M /0.07M_\sun$)  or $\sim$0.08($M /0.03M_\sun$) AU, respectively,
 as seen in Figure \ref{fig:a_Mp}.
The effective temperature of the planet when it was formed, based on the host mass and the planet-host separation, 
is  $\sim 26$K,   $\sim 13$K or $\sim 7$K, the coldest planet found to date apart from those planets found by the 
direct imaging method, which can presently only find heavy gas-giant planets of more than a few Jupiter masses.
The effective temperature of these heavy gas-giants are a few hundreds K or higher due to their internal heat.
In either interpretation, planet MOA-2013-BLG-605Lb is orbiting around one of the least massive objects found to date at a 
very wide separation. The planet is the coldest exoplanet discovered so far. 
This is the first observed example of a Neptune-like exoplanet
in terms of mass and temperature
or a super-Earth with Neptune-like temperature,
which are important factors in any planetary formation theory.

The probability of detecting such wide separation low mass planets is very low, even 
by microlensing. The probability 
of a source crossing the planetary caustic is proportional to the size of the planetary 
caustic, $w_{\rm c}\sim 4q^{1/2}s^{-2}\sim0.01$ \citep{Han2006}, divided by half the circle 
with radius of separation $s$, i.e.,
$P\sim 4/\pi q^{1/2}s^{-3}\sim1\times 10^{-3} \,(s=2.4)$. 
It is an order of magnitude smaller than planets at $s\sim 1$, where $\sim10$ planets with Neptune-mass or less 
have been found by microlensing.
This may imply that such low-mass planets with masses less than that of Neptune at $a\simeq 10\, a_{\rm snow}$ are as common as
low-mass planets at a few times of the snow line \citep{sumi2010,Gould2010,cassan2012}.

This conclusion may challenge the standard core accretion model and other formation models \citep{ida04}
which predict few low-mass planets with Neptune-like orbit at $>$10 $a_{\rm snow}$.
More accurate measurements of the abundance and distribution of such low-mass ice planets are very important
in the study of the formation of Neptune and in the study of planet formation mechanisms in general.
The microlensing exoplanet search by NASA's WFIRST satellite is expected to detect hundreds of 
low mass planets with Neptune-like orbit and will constrain further planetary formation models.

\acknowledgments
TS acknowledges the financial support from the JSPS, JSPS23103002,JSPS24253004 and JSPS26247023.
The MOA project is supported by the grant JSPS25103508 and 23340064.
The OGLE project has received funding from the National Science Centre,
Poland, grant MAESTRO 2014/14/A/ST9/00121 to AU.
DPB acknowledges support from NSF grants AST-1009621 and AST-1211875,
as well as NASA grants NNX12AF54G and NNX13AF64G.
Work by IAB and PY was supported by the Marsden Fund of the Royal Society of New Zealand, contract no. MAU1104.
NJR is a Royal Society of New Zealand Rutherford Discovery Fellow.
AS, ML and MD acknowledge support from the Royal Society of New Zealand. AS is a University of Auckland Doctoral Scholar.
AG was supported by NSF grant AST 1103471 and NASA grant NNX12AB99G.
J.P.B., S.B., J.B.M. gratefully acknowledges support from ESO's DGDF 2014. 
JPB \& JB acknowledge the support of the Programme National de Plan\'{e}tologie, CNRS, and from PERSU Sorbonne Universit\'{e}.
Work by C.H. was supported by Creative Research  Initiative  Program
(2009-0081561) of National Research Foundation of Korea. 

\clearpage


\begin{deluxetable}{lccrrr}
\tabletypesize{\footnotesize}
\tabletypesize{\scriptsize}
\tablecaption{Datasets used in the analysis \label{tbl:dataset}}
\tablewidth{0pt}
\tablehead{
\colhead{Dataset}   & \colhead{Telescope} & \colhead{Filter}  & \colhead{$N_{\rm data}$}        & \colhead{$k$}     & \colhead{$e_{\rm min}$}   
}
\startdata
 MOA-Red         &  MOA-II 1.8\,m   & $R+I$  &  9675 &   1.092 &  0.0127  \\
 OGLE-I            &  Warsaw 1.3\,m    & $I$  &  5514  &   1.387 &  0.0109 \\
 OGLE-V           & Warsaw 1.3\,m    & $V$ &    64    &   1.571 &  0.0 \\
 CTIO-I             & SMARTS-CTIO 1.3\,m   & $I$   &   15    &   1.0           &  0.0 \\
 CTIO-V            & SMARTS-CTIO 1.3\,m   & $V$  &   15    &   1.0           &  0.0 \\
 CTIO-H            & SMARTS-CTIO 1.3\,m   & $H$   &   149    &   1.0           &  0.0 \\
 \enddata
\tablecomments{
$N_{\rm data}$ is the number of data points used in the analysis.
$k$ and $e_{\rm min}$ are the error scaling parameters (see details in the text).
 }
\end{deluxetable}
\clearpage

\begin{deluxetable}{lrrrrr}
\tabletypesize{\footnotesize}
\tabletypesize{\scriptsize}
\tablecaption{Model Parameters.  Standard and Parallax-only models. \label{tbl:param}}
\tablewidth{0pt}
\tablehead{
\colhead{Parameter}         & \colhead{Standard}        & \colhead{P$_{+-}$}     & \colhead{P$_{-+}$}  & \colhead{P$_{++}$}     & \colhead{P$_{--}$}  
}
\startdata
 $t_{\rm 0}$ (HJD$'$)    &  6573.056 &  6573.050 &  6573.050 &  6573.051 &  6573.052 \\
                         &     0.008 &     0.009 &     0.009 &     0.009 &     0.010 \\
 $t_{\rm E}$ (days)      &     20.47 &     19.93 &     20.04 &     20.10 &     20.15 \\
                         &      0.13 &      0.32 &      0.29 &      0.32 &      0.28 \\
 $u_{\rm 0}$ ($10^{-2}$) &     7.563 &     7.932 &    -7.874 &     7.907 &    -7.890 \\
                         &     0.099 &     0.186 &     0.175 &     0.185 &     0.164 \\
 $q$ ($10^{-4}$)         &     2.762 &     3.460 &     3.431 &     3.530 &     3.611 \\
                         &     0.105 &     0.211 &     0.204 &     0.202 &     0.215 \\
\enddata
\tablecomments{
HJD$' = $HJD-2450000. The first subscript of model P, ``$_+$'' and ``$_-$'' indicate the models with  $u_0>0$ and $u_0<0$, respectively. 
The second subscript indicates the sign of impact parameter to the secondary lens, i.e, ``$_{++}$'' and ``$_{--}$'' mean
the source passes on the same side to the host and planet.  
The 1$\sigma$ error is given below each parameter.
 $\chi^2_{\rm lc}$ is the  $\chi^2$ from the light curve alone. $\Delta\chi^2_{\rm kep}$,   $\Delta\chi^2_{\rm kin}$ and  $\Delta\chi^2_{\rm gal}$ are
  $\chi^2$ penalty due to the Keplerian, kinematic and the Galactic priors. 
Note that $\pi_{\rm E}$ is not a fit parameter.
 }
\end{deluxetable}
\clearpage

\begin{deluxetable}{lrrrrrr}
\tiny
\tabletypesize{\scriptsize}
\tablecaption{Model Parameters.  Linear orbit (L).\label{tbl:param1}} 
\tablewidth{0pt}
\tablehead{
\colhead{Parameter}      
& \colhead{P$_{+-}$L}     & \colhead{P$_{-+}$L}   & \colhead{P$_{++}$L}     & \colhead{P$_{--}$L} 
& \colhead{P$_{++}$L$'$}     & \colhead{P$_{--}$L$'$} 
}
\startdata
 $t_{\rm 0}$ (HJD$'$)    &  6573.051 &  6573.051 &  6573.055 &  6573.055 &  6573.055 &  6573.055 \\
                         &     0.006 &     0.009 &     0.010 &     0.010 &     0.010 &     0.010 \\
 $t_{\rm E}$ (days)      &     19.95 &     19.90 &     21.33 &     21.24 &     21.25 &     21.22 \\
                         &      0.32 &      0.22 &      0.31 &      0.33 &      0.35 &      0.34 \\
 $u_{\rm 0}$ ($10^{-2}$) &     7.869 &    -7.894 &     7.418 &    -7.452 &     7.455 &    -7.449 \\
                         &     0.187 &     0.128 &     0.154 &     0.169 &     0.169 &     0.161 \\
 $q$ ($10^{-4}$)         &     3.614 &     3.422 &     4.183 &     4.913 &     2.805 &     2.846 \\
                         &     0.201 &     0.260 &     0.355 &     0.474 &     0.491 &     0.220 \\
\enddata
\tablecomments{
 Notation is ame as Table \ref{tbl:param}.  The solutions P$_{\pm\pm}$L$'$ does not exist with the full  Keplerian orbit.
 }
\end{deluxetable}
\clearpage

\begin{deluxetable}{lrrrr}
\tiny
\tabletypesize{\scriptsize}
\tablecaption{Model Parameters. Full keplerian orbit (K).\label{tbl:param5}} 
\tablewidth{0pt}
\tablehead{
\colhead{Parameter}      
& \colhead{P$_{+-}$K}     & \colhead{P$_{-+}$K}   & \colhead{P$_{++}$K}     & \colhead{P$_{--}$K} 
}
\startdata
 $t_{\rm 0}$ (HJD$'$)    &  6573.051 &  6573.051 &  6573.054 &  6573.055 \\
                         &     0.010 &     0.008 &     0.009 &     0.008 \\
 $t_{\rm E}$ (days)      &     19.98 &     19.91 &     21.03 &     21.14 \\
                         &      0.22 &      0.27 &      0.22 &      0.18 \\
 $u_{\rm 0}$ ($10^{-2}$) &     7.856 &    -7.886 &     7.553 &    -7.488 \\
                         &     0.134 &     0.157 &     0.119 &     0.085 \\
 $q$ ($10^{-4}$)         &     3.789 &     3.663 &     4.216 &     5.197 \\
                         &     0.196 &     0.211 &     0.335 &     0.317 \\
\enddata
\tablecomments{
  Notation is the same as for Table \ref{tbl:param}. 
 }
\end{deluxetable}
\clearpage

\begin{deluxetable}{lrrrrrrr}
\tiny
\tabletypesize{\scriptsize}
\tablecaption{Model Parameters.  Full keplerian orbit with keplerian prior (K$_{\rm p}$).\label{tbl:param6}} 
\tablewidth{0pt}
\tablehead{
\colhead{Parameter}
& \multicolumn{2}{c}{M-dwarf} & \multicolumn{2}{c}{high-mass brown dwarf} & \multicolumn{2}{c}{low-mass brown dwarf}\\
\cmidrule(r){2-3}\cmidrule(l){4-5}\cmidrule(l){6-7}      
& \colhead{P$_{+-}$K$_{\rm p}$}     & \colhead{P$_{-+}$K$_{\rm p}$}  
& \colhead{P$_{+-}$K$_{\rm p}$}     & \colhead{P$_{-+}$K$_{\rm p}$}   & \colhead{P$_{++}$K$_{\rm p}$}     & \colhead{P$_{--}$K$_{\rm p}$} 
}
\startdata
 $t_{\rm 0}$ (HJD$'$)    &  6573.051 &  6573.052 &  6573.050 &  6573.049 &  6573.053 &  6573.053 \\
                         &     0.009 &     0.009 &     0.009 &     0.009 &     0.008 &     0.009 \\
 $t_{\rm E}$ (days)      &     19.93 &     19.93 &     20.01 &     20.08 &     20.38 &     20.63 \\
                         &      0.27 &      0.19 &      0.31 &      0.26 &      0.19 &      0.16 \\
 $u_{\rm 0}$ ($10^{-2}$) &     7.942 &    -7.944 &     7.862 &    -7.816 &     7.816 &    -7.687 \\
                         &     0.165 &     0.120 &     0.177 &     0.150 &     0.134 &     0.080 \\
 $q$ ($10^{-4}$)         &     3.420 &     3.330 &     3.436 &     3.401 &     3.838 &     4.010 \\
                         &     0.165 &     0.118 &     0.132 &     0.160 &     0.185 &     0.155 \\
\enddata
\tablecomments{
  Notation is the same as for Table \ref{tbl:param}.
 }
\end{deluxetable}
\clearpage

\begin{deluxetable}{lrrrrrrr}
\tiny
\tabletypesize{\scriptsize}
\tablecaption{Model Parameters.  Kepler prior + kinematic constraint (K$_{\rm pk}$)\label{tbl:param16}} 
\tablewidth{0pt}
\tablehead{
\colhead{Parameter}      
& \multicolumn{2}{c}{M-dwarf} & \multicolumn{2}{c}{high-mass brown dwarf} & \multicolumn{2}{c}{low-mass brown dwarf}\\
\cmidrule(r){2-3}\cmidrule(l){4-5}\cmidrule(l){6-7}  
& \colhead{P$_{+-}$K$_{\rm pk}$}     & \colhead{P$_{-+}$K$_{\rm pk}$}  
& \colhead{P$_{+-}$K$_{\rm pk}$}     & \colhead{P$_{-+}$K$_{\rm pk}$}   & \colhead{P$_{++}$K$_{\rm pk}$}     & \colhead{P$_{--}$K$_{\rm pk}$} 
}
\startdata
 $t_{\rm 0}$ (HJD$'$)    &  6573.047 &  6573.054 &  6573.050 &  6573.051 &  6573.052 &  6573.054 \\
                         &     0.009 &     0.009 &     0.009 &     0.008 &     0.009 &     0.010 \\
 $t_{\rm E}$ (days)      &     20.20 &     19.83 &     19.96 &     19.93 &     20.41 &     20.52 \\
                         &      0.16 &      0.23 &      0.10 &      0.24 &      0.32 &      0.28 \\
 $u_{\rm 0}$ ($10^{-2}$) &     7.797 &    -7.960 &     7.890 &    -7.905 &     7.804 &    -7.749 \\
                         &     0.119 &     0.147 &     0.072 &     0.153 &     0.187 &     0.154 \\
 $q$ ($10^{-4}$)         &     3.526 &     3.124 &     3.557 &     3.400 &     3.812 &     4.054 \\
                         &     0.176 &     0.120 &     0.131 &     0.158 &     0.189 &     0.279 \\
\enddata
\tablecomments{
  Notation is the same as for Table \ref{tbl:param}.
 }
\end{deluxetable}
\clearpage

\begin{deluxetable}{lrrrrrrrr}
\tiny
\tabletypesize{\scriptsize}
\tablecaption{Model Parameters.  Keplerian prior + kinematic+Galactic prior (K$_{\rm pkg}$)\label{tbl:param26}} 
\tablewidth{0pt}
\tablehead{
\colhead{Parameter}
& \multicolumn{2}{c}{M-dwarf} & \multicolumn{4}{c}{high-mass brown dwarf} & \multicolumn{2}{c}{low-mass brown dwarf}\\
\cmidrule(r){2-3}\cmidrule(l){4-7}\cmidrule(l){8-9} 
& \colhead{P$_{+-}$K$_{\rm pkg}$}     & \colhead{P$_{-+}$K$_{\rm pkg}$}  
& \colhead{P$_{+-}$K$_{\rm pkg}$}     & \colhead{P$_{-+}$K$_{\rm pkg}$}   & \colhead{P$_{++}$K$_{\rm pkg}$}     & \colhead{P$_{--}$K$_{\rm pkg}$} 
& \colhead{P$_{++}$K$_{\rm pkg}$}     & \colhead{P$_{--}$K$_{\rm pkg}$} 
}
\startdata
 $t_{\rm 0}$ (HJD$'$)    &  6573.048 &  6573.055 &  6573.051 &  6573.052 &  6573.052 &  6573.051 &  6573.052 &  6573.052 \\
                         &     0.009 &     0.009 &     0.009 &     0.009 &     0.009 &     0.009 &     0.008 &     0.008 \\
 $t_{\rm E}$ (days)      &     20.14 &     19.70 &     20.01 &     19.88 &     19.96 &     20.09 &     20.43 &     20.64 \\
                         &      0.21 &      0.25 &      0.29 &      0.28 &      0.26 &      0.22 &      0.30 &      0.25 \\
 $u_{\rm 0}$ ($10^{-2}$) &     7.796 &    -8.042 &     7.860 &    -7.919 &     7.956 &    -7.897 &     7.778 &    -7.672 \\
                         &     0.104 &     0.162 &     0.172 &     0.165 &     0.159 &     0.130 &     0.160 &     0.140 \\
 $q$ ($10^{-4}$)         &     3.514 &     3.137 &     3.436 &     3.364 &     3.379 &     3.307 &     3.754 &     3.969 \\
                         &     0.070 &     0.082 &     0.146 &     0.155 &     0.203 &     0.196 &     0.240 &     0.274 \\
\enddata
\tablecomments{
  Notation is the same as for Table \ref{tbl:param}.
 }
\end{deluxetable}
\clearpage
\begin{deluxetable}{lcccc}
\tabletypesize{\footnotesize}
\tablecaption{Lens physical parameters  (Parallax only).\label{tbl:physical}}
\tablewidth{0pt}
\tablehead{
\colhead{Parameter}        & \colhead{P$_{+-}$}     & \colhead{P$_{-+}$}  & \colhead{P$_{++}$}     & \colhead{P$_{--}$}   
}
\startdata
 $\theta_{\rm E}$ (mas)              &          0.482$\pm$0.036    &          0.486$\pm$0.036    &          0.478$\pm$0.036    &          0.474$\pm$0.035    \\
 $\mu_{\rm geo}$   (mas yr$^{-1}$)   &           8.61$\pm$ 0.67    &           8.63$\pm$ 0.67    &           8.47$\pm$ 0.66    &           8.38$\pm$ 0.64    \\
 $\mu_{\rm hel,N}$   (mas yr$^{-1}$) &          -6.83$\pm$ 0.52    &           6.19$\pm$ 0.49    &           7.98$\pm$ 0.65    &          -8.54$\pm$ 0.63    \\
 $\mu_{\rm hel,E}$   (mas yr$^{-1}$) &          -5.73$\pm$ 0.42    &          -6.22$\pm$ 0.46    &          -2.51$\pm$ 0.12    &          -2.75$\pm$ 0.14    \\
 $D_{\rm l}$ (kpc)                   &   3.14$_{ -0.48}^{+  0.44}$ &   3.22$_{ -0.49}^{+  0.43}$ &   1.50$_{ -0.10}^{+  0.15}$ &   1.47$_{ -0.10}^{+  0.14}$ \\
 $M_{\rm h}$ ($M_\sun$)              &  0.147$_{-0.037}^{+ 0.031}$ &  0.156$_{-0.040}^{+ 0.030}$ &  0.052$_{-0.005}^{+ 0.008}$ &  0.050$_{-0.005}^{+ 0.007}$ \\
 $M_{\rm p}$ ($M_\earth$)            &  16.99$_{ -3.59}^{+  2.99}$ &  17.83$_{ -3.87}^{+  2.93}$ &   6.09$_{ -0.45}^{+  0.83}$ &   5.98$_{ -0.44}^{+  0.75}$ \\
 $a_\perp$ (AU)                      &   3.62$_{ -0.60}^{+  0.51}$ &   3.73$_{ -0.61}^{+  0.50}$ &   1.71$_{ -0.10}^{+  0.15}$ &   1.67$_{ -0.11}^{+  0.13}$ \\
\enddata
\tablecomments{
$a_{\rm exp}=\sqrt{3/2}a_\perp$ is an expected 3D separation for the measured projected separation, $a_\perp$.
$a_{\rm snow}=2.7(M_h/M_\sun$) is the snow-line.  $T_{\rm eff, exp}$ is the effective temperature
of the planet when it was formed based on the  $M_{\rm h}$ and  $a_{\rm exp}$.
}
\end{deluxetable}
\clearpage

\begin{deluxetable}{lcccc}
\tabletypesize{\footnotesize}
\tablecaption{Lens physical parameters.  Linear orbit (L)\label{tbl:physical1}}  
\tablewidth{0pt}
\tablehead{
\colhead{Parameter} 
& \colhead{P$_{+-}$L}     & \colhead{P$_{-+}$L}   & \colhead{P$_{++}$L}     & \colhead{P$_{--}$L} 
}
\startdata
 $\theta_{\rm E}$ (mas)              &          0.475$\pm$0.034    &          0.485$\pm$0.039    &          0.460$\pm$0.037    &          0.424$\pm$0.035    \\
 $\mu_{\rm geo}$   (mas yr$^{-1}$)   &           8.48$\pm$ 0.63    &           8.69$\pm$ 0.72    &           7.69$\pm$ 0.64    &           7.11$\pm$ 0.61    \\
 $\mu_{\rm hel,N}$   (mas yr$^{-1}$) &          -8.83$\pm$ 0.63    &           8.17$\pm$ 0.72    &           6.61$\pm$ 0.64    &          -7.99$\pm$ 0.61    \\
 $\mu_{\rm hel,E}$   (mas yr$^{-1}$) &          -2.59$\pm$ 0.10    &          -2.37$\pm$ 0.10    &          -2.84$\pm$ 0.01    &          -2.77$\pm$ 0.03    \\
 $D_{\rm l}$ (kpc)                   &   1.16$_{ -0.17}^{+  0.12}$ &   1.22$_{ -0.04}^{+  0.40}$ &   0.54$_{ -0.05}^{+  0.06}$ &   0.64$_{ -0.05}^{+  0.08}$ \\
 $M_{\rm h}$ ($M_\sun$)              &  0.038$_{-0.006}^{+ 0.004}$ &  0.042$_{-0.003}^{+ 0.015}$ &  0.015$_{-0.000}^{+ 0.002}$ &  0.015$_{-0.001}^{+ 0.005}$ \\
 $M_{\rm p}$ ($M_\earth$)            &   4.54$_{ -0.74}^{+  0.55}$ &   4.74$_{ -0.21}^{+  1.79}$ &   2.11$_{ -0.23}^{+  0.23}$ &   2.53$_{ -0.24}^{+  0.36}$ \\
 $a_\perp$ (AU)                      &   1.31$_{ -0.19}^{+  0.11}$ &   1.39$_{ -0.05}^{+  0.43}$ &   0.58$_{ -0.05}^{+  0.05}$ &   0.67$_{ -0.04}^{+  0.11}$ \\
\enddata
\tablecomments{
 Notation is the same as for Table \ref{tbl:physical}.
}
\end{deluxetable}
\clearpage

\begin{deluxetable}{lcccc}
\tabletypesize{\footnotesize}
\tablecaption{Lens physical parameters. Keplerian orbit (K)\label{tbl:physical5}} 
\tablewidth{0pt}
\tablehead{
\colhead{Parameter}   
& \colhead{P$_{+-}$K}     & \colhead{P$_{-+}$K}   & \colhead{P$_{++}$K}     & \colhead{P$_{--}$K} 
}
\startdata
 $\theta_{\rm E}$ (mas)              &          0.464$\pm$0.032    &          0.469$\pm$0.034    &          0.454$\pm$0.035    &          0.410$\pm$0.031    \\
 $\mu_{\rm geo}$   (mas yr$^{-1}$)   &           8.28$\pm$ 0.60    &           8.40$\pm$ 0.63    &           7.70$\pm$ 0.61    &           6.91$\pm$ 0.54    \\
 $\mu_{\rm hel,N}$   (mas yr$^{-1}$) &          -8.61$\pm$ 0.59    &           7.90$\pm$ 0.62    &           6.73$\pm$ 0.61    &          -7.76$\pm$ 0.54    \\
 $\mu_{\rm hel,E}$   (mas yr$^{-1}$) &          -2.59$\pm$ 0.10    &          -2.31$\pm$ 0.09    &          -2.60$\pm$ 0.01    &          -2.62$\pm$ 0.02    \\
 $D_{\rm l}$ (kpc)                   &   1.20$_{ -0.03}^{+  0.39}$ &   1.26$_{ -0.08}^{+  0.27}$ &   0.60$_{ -0.05}^{+  0.06}$ &   0.67$_{ -0.05}^{+  0.07}$ \\
 $M_{\rm h}$ ($M_\sun$)              &  0.037$_{-0.001}^{+ 0.014}$ &  0.040$_{-0.004}^{+ 0.010}$ &  0.016$_{-0.001}^{+ 0.003}$ &  0.015$_{-0.001}^{+ 0.002}$ \\
 $M_{\rm p}$ ($M_\earth$)            &   4.70$_{ -0.16}^{+  1.69}$ &   4.93$_{ -0.23}^{+  1.37}$ &   2.31$_{ -0.16}^{+  0.28}$ &   2.60$_{ -0.17}^{+  0.35}$ \\
 $a_\perp$ (AU)                      &   1.35$_{ -0.03}^{+  0.40}$ &   1.43$_{ -0.08}^{+  0.28}$ &   0.65$_{ -0.04}^{+  0.04}$ &   0.69$_{ -0.04}^{+  0.05}$ \\
\enddata
\tablecomments{
 Notation is the same as for Table \ref{tbl:physical}, except $T_{\rm eff, kep}$ is based on $a_{\rm kep}$.
When the best fit is larger or smaller than the 68\% confidence interval of MCMC chains, 
the upper or lower limit is designated as ``$+0.0$'' or ``$-0.0$'', respectively.
}
\end{deluxetable}
\clearpage

\begin{deluxetable}{lcccccc}
\tabletypesize{\footnotesize}
\tablecaption{Lens physical parameters.   Keplerian orbit with Keplerian prior (K$_{\rm p}$)\label{tbl:physical6}} 
\tablewidth{0pt}
\tablehead{
\colhead{Parameter} 
& \multicolumn{2}{c}{M-dwarf} & \multicolumn{2}{c}{high-mass brown dwarf} & \multicolumn{2}{c}{low-mass brown dwarf}\\
\cmidrule(r){2-3}\cmidrule(l){4-5}\cmidrule(l){6-7}      
& \colhead{P$_{+-}$K$_{\rm p}$}     & \colhead{P$_{-+}$K$_{\rm p}$}  
& \colhead{P$_{+-}$K$_{\rm p}$}     & \colhead{P$_{-+}$K$_{\rm p}$}   & \colhead{P$_{++}$K$_{\rm p}$}     & \colhead{P$_{--}$K$_{\rm p}$} 
}
\startdata
 $\theta_{\rm E}$ (mas)              &          0.482$\pm$0.035    &          0.491$\pm$0.034    &          0.487$\pm$0.034    &          0.489$\pm$0.035    &          0.462$\pm$0.033    &          0.458$\pm$0.031    \\
 $\mu_{\rm geo}$   (mas yr$^{-1}$)   &           8.62$\pm$ 0.65    &           8.78$\pm$ 0.63    &           8.66$\pm$ 0.63    &           8.68$\pm$ 0.64    &           8.07$\pm$ 0.59    &           7.90$\pm$ 0.56    \\
 $\mu_{\rm hel,N}$   (mas yr$^{-1}$) &          -0.27$\pm$ 0.02    &           1.13$\pm$ 0.09    &          -8.47$\pm$ 0.60    &           8.08$\pm$ 0.62    &           7.41$\pm$ 0.59    &          -8.51$\pm$ 0.56    \\
 $\mu_{\rm hel,E}$   (mas yr$^{-1}$) &          -8.81$\pm$ 0.65    &          -8.88$\pm$ 0.63    &          -3.43$\pm$ 0.20    &          -3.14$\pm$ 0.18    &          -2.19$\pm$ 0.03    &          -2.57$\pm$ 0.05    \\
 $D_{\rm l}$ (kpc)                   &   4.22$_{ -0.97}^{+  0.63}$ &   4.31$_{ -0.78}^{+  0.80}$ &   1.93$_{ -0.23}^{+  0.40}$ &   1.95$_{ -0.11}^{+  0.45}$ &   0.86$_{ -0.06}^{+  0.07}$ &   0.86$_{ -0.09}^{+  0.04}$ \\
 $M_{\rm h}$ ($M_\sun$)              &  0.256$_{-0.094}^{+ 0.061}$ &  0.276$_{-0.083}^{+ 0.095}$ &  0.074$_{-0.012}^{+ 0.020}$ &  0.076$_{-0.007}^{+ 0.018}$ &  0.025$_{-0.002}^{+ 0.003}$ &  0.025$_{-0.003}^{+ 0.002}$ \\
 $M_{\rm p}$ ($M_\earth$)            &  29.13$_{ -9.98}^{+  6.53}$ &  30.68$_{ -8.71}^{+ 10.12}$ &   8.45$_{ -1.18}^{+  2.37}$ &   8.62$_{ -0.44}^{+  2.61}$ &   3.21$_{ -0.27}^{+  0.30}$ &   3.33$_{ -0.40}^{+  0.19}$ \\
 $a_\perp$ (AU)                      &   4.86$_{ -1.12}^{+  0.75}$ &   5.02$_{ -0.92}^{+  0.94}$ &   2.23$_{ -0.26}^{+  0.42}$ &   2.27$_{ -0.13}^{+  0.43}$ &   0.95$_{ -0.04}^{+  0.04}$ &   0.95$_{ -0.09}^{+  0.01}$ \\
\enddata
\tablecomments{
 Notation is the same as for Table \ref{tbl:physical5}.
}
\end{deluxetable}
\clearpage
 
\begin{deluxetable}{lcccccc}
\tabletypesize{\footnotesize}
\tablecaption{Lens physical parameters.  Keplerian prior + kinematic constraint (K$_{\rm pk}$)\label{tbl:physical16}} 
\tablewidth{0pt}
\tablehead{
\colhead{Parameter} 
& \multicolumn{2}{c}{M-dwarf} & \multicolumn{2}{c}{high-mass brown dwarf} & \multicolumn{2}{c}{low-mass brown dwarf}\\
\cmidrule(r){2-3}\cmidrule(l){4-5}\cmidrule(l){6-7}  
& \colhead{P$_{+-}$K$_{\rm pk}$}     & \colhead{P$_{-+}$K$_{\rm pk}$}  
& \colhead{P$_{+-}$K$_{\rm pk}$}     & \colhead{P$_{-+}$K$_{\rm pk}$}   & \colhead{P$_{++}$K$_{\rm pk}$}     & \colhead{P$_{--}$K$_{\rm pk}$} 
}
\startdata
 $\theta_{\rm E}$ (mas)              &          0.481$\pm$0.035    &          0.503$\pm$0.034    &          0.477$\pm$0.033    &          0.486$\pm$0.035    &          0.464$\pm$0.033    &          0.452$\pm$0.034    \\
 $\mu_{\rm geo}$   (mas yr$^{-1}$)   &           8.48$\pm$ 0.63    &           9.03$\pm$ 0.63    &           8.51$\pm$ 0.61    &           8.68$\pm$ 0.65    &           8.11$\pm$ 0.61    &           7.86$\pm$ 0.61    \\
 $\mu_{\rm hel,N}$   (mas yr$^{-1}$) &           1.16$\pm$ 0.09    &           8.26$\pm$ 0.59    &          -8.45$\pm$ 0.59    &           8.19$\pm$ 0.63    &           7.44$\pm$ 0.61    &          -8.48$\pm$ 0.61    \\
 $\mu_{\rm hel,E}$   (mas yr$^{-1}$) &          -8.65$\pm$ 0.63    &          -3.69$\pm$ 0.24    &          -3.14$\pm$ 0.17    &          -2.72$\pm$ 0.15    &          -2.20$\pm$ 0.03    &          -2.47$\pm$ 0.05    \\
 $D_{\rm l}$ (kpc)                   &   3.59$_{ -0.39}^{+  0.60}$ &   3.55$_{ -0.78}^{+  0.20}$ &   1.76$_{ -0.15}^{+  0.19}$ &   1.92$_{ -0.12}^{+  0.28}$ &   0.86$_{ -0.06}^{+  0.12}$ &   0.85$_{ -0.08}^{+  0.08}$ \\
 $M_{\rm h}$ ($M_\sun$)              &  0.185$_{-0.034}^{+ 0.054}$ &  0.198$_{-0.065}^{+ 0.007}$ &  0.063$_{-0.006}^{+ 0.006}$ &  0.073$_{-0.006}^{+ 0.014}$ &  0.025$_{-0.003}^{+ 0.004}$ &  0.024$_{-0.003}^{+ 0.004}$ \\
 $M_{\rm p}$ ($M_\earth$)            &  21.77$_{ -3.34}^{+  5.31}$ &  20.58$_{ -6.66}^{+  0.56}$ &   7.47$_{ -0.77}^{+  0.87}$ &   8.28$_{ -0.67}^{+  1.35}$ &   3.23$_{ -0.29}^{+  0.55}$ &   3.24$_{ -0.30}^{+  0.38}$ \\
 $a_\perp$ (AU)                      &   4.13$_{ -0.48}^{+  0.74}$ &   4.18$_{ -0.96}^{+  0.23}$ &   2.01$_{ -0.14}^{+  0.15}$ &   2.22$_{ -0.12}^{+  0.29}$ &   0.95$_{ -0.08}^{+  0.11}$ &   0.93$_{ -0.08}^{+  0.08}$ \\
\enddata
\tablecomments{
 Notation is the same as for Table \ref{tbl:physical5}.
}
\end{deluxetable}
\clearpage

\begin{deluxetable}{lcccccccc}
\rotate
\tabletypesize{\footnotesize}
\tabletypesize{\scriptsize}
\tablecaption{Lens physical parameters. Keplerian prior + kinematic + Galactic density prior (K$_{\rm pkg}$) \label{tbl:physical26}} 
\tablewidth{0pt}
\tablehead{ 
\colhead{Parameter}
& \multicolumn{2}{c}{M-dwarf} & \multicolumn{4}{c}{high-mass brown dwarf} & \multicolumn{2}{c}{low-mass brown dwarf}\\
\cmidrule(r){2-3}\cmidrule(l){4-7}\cmidrule(l){8-9} 
& \colhead{P$_{+-}$K$_{\rm pkg}$}     & \colhead{P$_{-+}$K$_{\rm pkg}$}  
& \colhead{P$_{+-}$K$_{\rm pkg}$}     & \colhead{P$_{-+}$K$_{\rm pkg}$}   & \colhead{P$_{++}$K$_{\rm pkg}$}     & \colhead{P$_{--}$K$_{\rm pkg}$} 
& \colhead{P$_{++}$K$_{\rm pkg}$}     & \colhead{P$_{--}$K$_{\rm pkg}$} 
}
\startdata
 $\theta_{\rm E}$ (mas)              &          0.482$\pm$0.033    &          0.498$\pm$0.034    &          0.487$\pm$0.035    &          0.487$\pm$0.034    &          0.485$\pm$0.036    &          0.493$\pm$0.036    &          0.469$\pm$0.035    &          0.461$\pm$0.035    \\
 $\mu_{\rm geo}$   (mas yr$^{-1}$)   &           8.53$\pm$ 0.60    &           9.01$\pm$ 0.64    &           8.66$\pm$ 0.65    &           8.72$\pm$ 0.63    &           8.65$\pm$ 0.67    &           8.74$\pm$ 0.66    &           8.17$\pm$ 0.63    &           7.96$\pm$ 0.62    \\
 $\mu_{\rm hel,N}$   (mas yr$^{-1}$) &           1.71$\pm$ 0.13    &           8.51$\pm$ 0.61    &          -8.47$\pm$ 0.61    &           8.28$\pm$ 0.62    &           8.22$\pm$ 0.66    &          -8.67$\pm$ 0.63    &           7.52$\pm$ 0.63    &          -8.56$\pm$ 0.62    \\
 $\mu_{\rm hel,E}$   (mas yr$^{-1}$) &          -8.59$\pm$ 0.59    &          -2.96$\pm$ 0.19    &          -3.42$\pm$ 0.20    &          -2.53$\pm$ 0.14    &          -2.47$\pm$ 0.14    &          -3.17$\pm$ 0.18    &          -2.16$\pm$ 0.03    &          -2.53$\pm$ 0.06    \\
 $D_{\rm l}$ (kpc)                   &   3.74$_{ -0.55}^{+  0.68}$ &   3.70$_{ -0.55}^{+  0.49}$ &   1.93$_{ -0.19}^{+  0.38}$ &   2.00$_{ -0.26}^{+  0.20}$ &   1.95$_{ -0.16}^{+  0.16}$ &   1.81$_{ -0.08}^{+  0.38}$ &   0.86$_{ -0.04}^{+  0.10}$ &   0.87$_{ -0.08}^{+  0.07}$ \\
 $M_{\rm h}$ ($M_\sun$)              &  0.201$_{-0.045}^{+ 0.053}$ &  0.210$_{-0.050}^{+ 0.033}$ &  0.074$_{-0.010}^{+ 0.022}$ &  0.077$_{-0.016}^{+ 0.009}$ &  0.074$_{-0.009}^{+ 0.007}$ &  0.070$_{-0.005}^{+ 0.023}$ &  0.026$_{-0.000}^{+ 0.005}$ &  0.026$_{-0.004}^{+ 0.003}$ \\
 $M_{\rm p}$ ($M_\earth$)            &  23.54$_{ -5.07}^{+  6.29}$ &  21.96$_{ -4.88}^{+  3.84}$ &   8.45$_{ -1.04}^{+  2.26}$ &   8.66$_{ -1.46}^{+  1.07}$ &   8.35$_{ -0.71}^{+  0.75}$ &   7.67$_{ -0.28}^{+  2.40}$ &   3.27$_{ -0.20}^{+  0.43}$ &   3.39$_{ -0.35}^{+  0.29}$ \\
\enddata
\tablecomments{
 Notation is the same as for Table \ref{tbl:physical5}.
}
\end{deluxetable}
\clearpage

\end{document}